\RequirePackage{amsmath}
\documentclass{lmcs}
\pdfoutput=1

\usepackage{lastpage}
\lmcsdoi{18}{3}{33}
\lmcsheading{}{\pageref{LastPage}}{}{}%
{Dec.~08,~2021}{Sep.~15,~2022}{}


\usepackage{mathpartir}
\usepackage{hyperref}
\usepackage[utf8]{inputenc}

\usepackage{cleveref}
\crefname{lstlisting}{listing}{listings}
\Crefname{lstlisting}{Listing}{Listings}
\crefname{lem}{lemma}{lemmas}
\crefname{prop}{proposition}{propositions}

\usepackage{mathpartir}
\usepackage{listings}
\usepackage{amssymb}
\let\vec\relax
\usepackage{MnSymbol} % \llangle, \rrangle

\usepackage{stmaryrd} % \llparenthesis, \rrparenthesis
\usepackage{subfigure}
\newcommand\Newinline{\colorbox{gray!15}}
\definecolor{newcolor}{gray}{0.95}
\newcommand\Syntax\textsf

\newcommand\Nchan{\alpha}            % fixed channel names
\newcommand\Nchanb\beta
\newcommand\Int{\Syntax{Int}}
\newcommand\Inp[1]{{?#1}.}
\newcommand\Outp[1]{{!#1}.}
\newcommand\Chan{\Syntax{Chan}~}
\newcommand\TUnit{\Syntax{Unit}}
\newcommand\Ap[1]{{[#1]}} % access point
\newcommand\End{\Syntax{End}}
\newcommand\Rownull{\emptyset}

\newcommand\TEnull{\emptyset}

\newcommand\sepimp{\mathrel{-\mkern-6mu*}}
\newcommand\linto{\sepimp}
\newcommand\lpair{\otimes}

\newcommand\Trans[1]{\llangle#1\rrangle}
\newcommand\TransE[2][\sigma]{\llangle#2\rrangle_{#1}}

\newcommand\Back[1]{\llparenthesis#1\rrparenthesis}
\newcommand\TOANF[1]{\llfloor#1\rrfloor}

\newcommand\Tagged[2]{{#2}_{#1}}

\newcommand\Erase[1]{|#1|}

\newcommand\KWFork{\Syntax{fork}}
\newcommand\KWReceive{\Syntax{receive}}
\newcommand\KWAccept{\Syntax{accept}}
\newcommand\KWRequest{\Syntax{request}}
\newcommand\KWSend{\Syntax{send}}
\newcommand\KWNew{\Syntax{new}}
\newcommand\KWClose{\Syntax{close}}

\newcommand\SendF{\KWSend\,}
\newcommand\Send[1]{\KWSend\,#1\,\Syntax{on}\,}
\newcommand\Receive{\KWReceive\,}
\newcommand\Accept{\KWAccept\,}
\newcommand\Request{\KWRequest\,}
\newcommand\Fork{\KWFork\,}
\newcommand\Fix{\Syntax{fix}\,}
\newcommand\New{\KWNew\,}
\newcommand\Close{\Syntax{close}\,}

\newcommand\Lam[1]{\lambda#1.}

\newcommand\UnitV{()}%{\Syntax{unit}}
\newcommand\Let[2]{\Syntax{let}\,#1=#2\,\Syntax{in}\,}
\newcommand\Record[1]{\{#1\}}
\newcommand\Remp{\Record{}}%{\Syntax{emp}
\newcommand\Rsingle[2]{\Record{#1=#2}}
\newcommand\Rconcat[1]{#1\cdot}
\newcommand\Rsplit[1]{#1.}%{\Syntax{split}\,#1,}
\newcommand\Rsplitmany[1]{#1.}%{\Syntax{split}^*\,#1,}

\newcommand\Hole\Box

\newcommand\ThreadOpen\langle
\newcommand\ThreadClose\rangle
\newcommand\Thread[1]{\ThreadOpen#1\ThreadClose}
\newcommand\Cnewchan[2]{(\nu #1#2)}
\newcommand\Cnewap[1]{(\nu #1)}

\newcommand\Fresh{\textup{\textrm{fresh}}\,}
\newcommand\Dual\overline
\newcommand\Dom[1]{\textup{\textrm{dom}} (#1)}
\newcommand\ReduceTo\rightarrow
\newcommand\GVEReducePlus[1][{}]{\stackrel{\smash{#1}}{\rightarrow}^+_e}
\newcommand\GVEReduceStar[1][{}]{\stackrel{\smash{#1}}{\rightarrow}^*_e}
\newcommand\GVEReduceTo[1][{}]{\stackrel{{#1}}{\rightarrow}_e}
\newcommand\VGREReduceTo[1][]{\stackrel{{#1}}{\Rightarrow}_e}%{\xRightarrow{#1}_e}%
\newcommand\VGREReducePlus[1][]{\stackrel{\smash{#1}}{\Rightarrow}^+_e}%{\xRightarrow{#1}_e}%
\newcommand\GVPReducePlus[1][{}]{\stackrel{\smash{#1}}{\rightarrow}^+_p}
\newcommand\GVPReduceTo[1][{}]{\stackrel{{#1}}{\rightarrow}_p}
\newcommand\VGRPReducePlus[1][{}]{\stackrel{\smash{#1}}{\Rightarrow}^+_p}%{\xRightarrow{#1}_p}%
\newcommand\VGRPReduceTo[1][{}]{\stackrel{{#1}}{\Rightarrow}_p}%{\xRightarrow{#1}_p}%

\newcommand\RLabel\ell          %reduction labels
\newcommand\ELabel\tau          %empty label

\newcommand\Pcong\equiv          %structural congruence of processes

\newcommand\Adorn[1]{{#1}^\bullet}

\newcommand\JVGRValue[3]{
  #1; #2 \mapsto #3
}
\newcommand\JVGRExpr[6]{
  #1; #2; #3 \mapsto #4; #5; #6
}
\newcommand\JVGRConfig[4]{
  #1;#2;#3 \mapsto #4
}

\newcommand\ruleVGRConst{
  \inferrule[C-Const]{}{
    \JVGRValue\Gamma\UnitV\TUnit
  }
}

\newcommand\ruleVGRChan{
  \inferrule[C-Chan]{}{
    \JVGRValue\Gamma{\gamma^p}{\Chan{\gamma^p}}
  }
}

\newcommand\ruleVGRVar{
  \inferrule[C-Var]{}{
    \JVGRValue{\Gamma,x:T}x T
  }
}

\newcommand\ruleVGRAbs{
  \inferrule[C-Abs]{
  \JVGRExpr{\Gamma, x:T} \Sigma e {\Sigma_1} U {\Sigma_2}
}{
  \JVGRValue\Gamma{\Lam xe}{(\Sigma; T \to U; \Sigma_1, \Sigma_2)}
}
}

\newcommand\ruleVGRReceiveD{
  \inferrule[C-ReceiveD]{
    \JVGRValue\Gamma v {\Chan\alpha}
  }{
    \JVGRExpr\Gamma{\Sigma, \alpha: \Inp{D}S}
    {\Receive v}
    \Sigma D {\alpha : S}
  }
}

\newcommand\ruleVGRReceiveS{
  \inferrule[C-ReceiveS]{
    \JVGRValue\Gamma v {\Chan\alpha} \\
    \Fresh d
  }{
    \JVGRExpr\Gamma{\Sigma, \alpha: \Inp{S'}S}
    {\Receive v}
    \Sigma {\Chan d} {d : S', \alpha : S}
  }
}

\newcommand\ruleVGRSendD{
  \inferrule[C-SendD]{
    \JVGRValue \Gamma v {D} \\
    \JVGRValue \Gamma {v'}{\Chan\alpha}
  }{
    \JVGRExpr \Gamma{\Sigma, \alpha: \Outp{D}{S}}
    {\Send v{v'}}
    \Sigma {\TUnit}{\alpha:S}
  }
}

\newcommand\ruleVGRSendS{
  \inferrule[C-SendS]{
    \JVGRValue \Gamma v {\Chan\beta} \\
    \JVGRValue \Gamma {v'}{\Chan\alpha}
  }{
    \JVGRExpr \Gamma{\Sigma, \alpha: \Outp{S'}{S}, \beta: S'}
    {\Send v{v'}}
    \Sigma {\TUnit}{\alpha:S}
  }
}

\newcommand\ruleVGRClose{
  \inferrule[C-Close]{
    \JVGRValue\Gamma v {\Chan\alpha}
  }{
    \JVGRExpr\Gamma{\Sigma, \alpha: \End}{\Close v}\Sigma \TUnit \emptyset
  }
}

\newcommand\ruleVGRAccept{
  \inferrule[C-Accept]{
    \JVGRValue\Gamma v{[S]} \\
    \Fresh c
  }{
    \JVGRExpr\Gamma\Sigma
    {\Accept v}
    \Sigma{\Chan c}{\{c:S\}}
  }
}

\newcommand\ruleVGRRequest{
  \inferrule[C-Request]{
    \JVGRValue\Gamma v{[S]} \\
    \Fresh c
  }{
    \JVGRExpr\Gamma\Sigma
    {\Request v}
    \Sigma{\Chan c}{\{c:\Dual S\}}
  }
}

\newcommand\ruleVGRVal{
  \inferrule[C-Val]{
    \JVGRValue\Gamma v T
  }{
    \JVGRExpr\Gamma\Sigma v \Sigma T \emptyset
  }
}

\newcommand\ruleVGRApp{
  \inferrule[C-App]{
    \JVGRValue\Gamma v {(\Sigma; T \to U; \Sigma')} \\
    \JVGRValue\Gamma{v'}{ T}
  }{
    \JVGRExpr\Gamma{\Sigma,\Sigma''}{v\,v'}{\Sigma''}U{\Sigma'}
  }
}

\newcommand\ruleVGRNew{
  \inferrule[C-New]{}{
    \JVGRExpr\Gamma\Sigma{\New S}\Sigma{\Ap S}\emptyset
  }
}

\newcommand\ruleVGRLet{
  \inferrule[C-Let]{
    \JVGRExpr\Gamma\Sigma e {\Sigma_1}{T_1}{\Sigma_1'} \\\\
    \JVGRExpr{\Gamma, x:T_1}{\Sigma_1,\Sigma_1'} t {\Sigma_2}{T_2}{\Sigma_2'}
  }{
    \JVGRExpr\Gamma\Sigma{\Let x e
      t}{\Sigma_1\cap\Sigma_2}{T_2}{\Sigma_1'\cap\Sigma_2, \Sigma_2'}
  }
}

\newcommand\ruleVGRForkFirstPremise{
  \JVGRExpr\Gamma \Sigma{t_1}{\Sigma_1}{T_1}{\{\}}
}
\newcommand\ruleVGRForkSecondPremise{
  \JVGRExpr\Gamma{\Sigma_1}{t_2}{\Sigma_2}{T_2}{\{\}}
}
\newcommand\ruleVGRFork{
  \inferrule[C-Fork-Orig]{
    \ruleVGRForkFirstPremise \\
    \ruleVGRForkSecondPremise
  }{
    \JVGRExpr\Gamma\Sigma{(\Fork t_1;t_2)}{\Sigma_2}{T_2}{\{\}}
  }
}

\newcommand\ruleVGRForkFirstPremiseRevised{
  \JVGRExpr\Gamma {\Sigma_1}{t_1}{\{\}}{T_1}{\{\}}
}
\newcommand\ruleVGRForkSecondPremiseRevised{
  \JVGRExpr\Gamma{\Sigma_2}{t_2}{\Sigma}{T_2}{\{\}}
}
\newcommand\ruleVGRForkRevised{
  \inferrule[C-Fork]{
    \ruleVGRForkFirstPremiseRevised \\
    \ruleVGRForkSecondPremiseRevised
  }{
    \JVGRExpr\Gamma{\Sigma_1,\Sigma_2}{(\Fork t_1;t_2)}{\Sigma}{T_2}{\{\}}
  }
}

\newcommand\ruleVGRThread{
  \inferrule[C-Thread]{
    \JVGRExpr{\vec x : [\vec S]}\Sigma t{\Sigma'} T {\{\}}
  }{
    \JVGRConfig{\vec x : [\vec S]}\Sigma{\Thread t}{\Sigma'}
  }
}

\newcommand\ruleVGRPar{
  \inferrule[C-Par]{
    \JVGRConfig\Gamma\Sigma{C_1}{\Sigma_1} \\
    \JVGRConfig\Gamma{\Sigma_1}{C_2}{\Sigma_2}
  }{
    \JVGRConfig\Gamma\Sigma{C_1 \| C_2 }{\Sigma_2}
  }
}

\newcommand\ruleVGRNewN{
  \inferrule[C-NewN]{
    \JVGRConfig{\Gamma, n:T}\Sigma C {\Sigma'}
  }{
    \JVGRConfig\Gamma\Sigma{(\nu n:T)C}{\Sigma'}
  }
}

\newcommand\ruleVGRNewB{
  \inferrule[C-NewB]{
    \JVGRConfig{\Gamma}{\Sigma, \gamma^+:S, \gamma^-:\Dual S} C
    {\Sigma'} \\
    {\gamma\notin\Gamma,\Sigma,\Sigma'}
  }{
    \JVGRConfig\Gamma\Sigma{(\nu \gamma)C}{\Sigma'}
  }
}

\newcommand\ruleVGRNewC{
  \inferrule[C-NewC]{
    \JVGRConfig{\Gamma}{\Sigma} C
    {\Sigma'} \\
    {\gamma\notin\Gamma,\Sigma,\Sigma'}
  }{
    \JVGRConfig\Gamma\Sigma{(\nu \gamma)C}{\Sigma'}
  }
}

\newcommand{\JGVExpr}[4][{}]{
  #2 \vdash#1 #3 : #4
}
\newcommand\JGVExprID{\JGVExpr[']}

\newcommand\Unr[1]{\textup{unr}\,#1}
\newcommand\GVsplit[3]{#1 = #2 + #3}

\newcommand\ruleGVvar[1][{}]{
  \inferrule[T-Var#1]{\Unr\Gamma}{
    \JGVExpr[#1]{\Gamma, x:t} x t
  }
}
\newcommand\ruleGVvarEFF{
  \inferrule[T-Var']{\Unr\Gamma}{
    \JGVExpr[']{\Gamma, x:t} x {t/\Sigma\mapsto\Sigma}
  }
}
\newcommand\ruleGVunit[1][{}]{
  \inferrule[T-UnitI#1]{\Unr\Gamma}{
    \JGVExpr[#1]\Gamma \UnitV \TUnit
  }
}
\newcommand\ruleGVunitEFF{
  \inferrule[T-Unit']{\Unr\Gamma}{
    \JGVExpr[']\Gamma \UnitV {\TUnit/\Sigma\mapsto\Sigma}
  }
}
\newcommand\ruleGVlamunr{
  \inferrule[T-LamU]{
    \Unr\Gamma \\
    \JGVExpr{\Gamma, x:t_2} e {t_1}
  }{
    \JGVExpr\Gamma{\Lam x e}{t_2 \to t_1}
  }
}
\newcommand\ruleGVlamunrEFF{
  \inferrule[T-LamU']{
    \Unr\Gamma \\
    \JGVExpr[']{\Gamma, x:t_2} e {t_1/\Sigma_0 \mapsto \Sigma_1}
  }{
    \JGVExpr[']\Gamma{\Lam x e}{t_2 \to^{\Sigma_0\mapsto\Sigma_1} t_1 / \Sigma\mapsto\Sigma}
  }
}

\newcommand\ruleGVlamlin{
  \inferrule[T-LamL]{
    \JGVExpr{\Gamma, x:t_2} e {t_1}
  }{
    \JGVExpr\Gamma{\Lam x e}{t_2 \linto t_1}
  }
}

\newcommand\ruleGVlamlinEFF{
  \inferrule[T-LamL']{
    \JGVExprID{\Gamma, x:t_2} e {t_1/\Sigma_0\mapsto\Sigma_1}
  }{
    \JGVExprID\Gamma{\Lam x e}{t_2 \linto^{\Sigma_0\mapsto\Sigma_1} {t_1/\Sigma\mapsto\Sigma}}
  }
}

\newcommand\ruleGVapp{
  \inferrule[T-AppU]{
    \JGVExpr{\Gamma_1}{e_1}{t_2 \to t_1} \\
    \JGVExpr{\Gamma_2}{e_2}{t_2}
  }{
    \JGVExpr{\Gamma_1+\Gamma_2}{e_1\,e_2}{t_1}
  }
}

\newcommand\ruleGVappLin{
  \inferrule[T-AppL]{
    \JGVExpr{\Gamma_1}{e_1}{t_2 {\linto} t_1} \\
    \JGVExpr{\Gamma_2}{e_2}{t_2}
  }{
    \JGVExpr{\Gamma_1+\Gamma_2}{e_1\,e_2}{t_1}
  }
}

\newcommand\ruleGVappEFF{
  \inferrule[T-App']{
    \GVsplit\Gamma{\Gamma_1}{\Gamma_2} \\
    \JGVExpr[']{\Gamma_1}{e_1}{t_2 {\linto}^{\Sigma_2\mapsto\Sigma_3}
      t_1 / \Sigma_0\mapsto\Sigma_1} \\
    \JGVExpr[']{\Gamma_2}{e_2}{t_2/\Sigma_1\mapsto\Sigma_2,\Sigma_2'}
  }{
    \JGVExpr[']\Gamma{e_1\,e_2}{t_1/\Sigma_0\mapsto\Sigma_3,\Sigma_2'}
  }
}
\newcommand\ruleGVpair{
  \inferrule[T-PairI]{
    \JGVExpr{\Gamma_1}{e_1}{t_1} \\
    \JGVExpr{\Gamma_2}{e_2}{t_2}
  }{
    \JGVExpr{\Gamma_1+\Gamma_2}{(e_1, e_2)}{t_1 \lpair t_2}
  }
}
\newcommand\ruleGVpairIEFF{
  \inferrule[T-PairI']{
    \GVsplit\Gamma{\Gamma_1}{\Gamma_2} \\
    \JGVExpr[']{\Gamma_1}{e_1}{t_1/\Sigma\mapsto\Sigma'} \\
    \JGVExpr[']{\Gamma_2}{e_2}{t_2/\Sigma'\mapsto\Sigma''}
  }{
    \JGVExpr[']\Gamma{(e_1, e_2)}{t_1 \lpair t_2/\Sigma\mapsto\Sigma''}
  }
}

\newcommand\ruleGVpairE{
  \inferrule[T-PairE]{
    \JGVExpr{\Gamma_1}{e_1}{t_1 \lpair t_2} \\
    \JGVExpr{\Gamma_2, x:t_1, y:t_2}{e_2}{t}
  }{
    \JGVExpr{\Gamma_1+\Gamma_2}{\Let{(x,y)}{e_1}{e_2}}{t}
  }
}

\newcommand\ruleGVunitE{
  \inferrule[T-UnitE]{
    \JGVExpr{\Gamma_1}{e_1}{\TUnit} \\
    \JGVExpr{\Gamma_2}{e_2}{t}
  }{
    \JGVExpr{\Gamma_1+\Gamma_2}{\Let{()}{e_1}{e_2}}{t}
  }
}

\newcommand\ruleGVpairEEFF{
  \inferrule[T-PairE']{
    \GVsplit\Gamma{\Gamma_1}{\Gamma_2} \\
    \JGVExpr{\Gamma_1}{e_1}{t_1 \lpair t_2/\Sigma\mapsto\Sigma'} \\
    \JGVExpr{\Gamma_2, x:t_1, y:t_2}{e_2}{t/\Sigma'\mapsto\Sigma''}
  }{
    \JGVExpr\Gamma{\Let{(x,y)}{e_1}{e_2}}{t/\Sigma\mapsto\Sigma''}
  }
}

\newcommand\ruleGVletEFF{
  \inferrule[T-Let']{
    \GVsplit\Gamma{\Gamma_1}{\Gamma_2} \\
    \JGVExpr{\Gamma_1}{e_1}{t_1/\Sigma\mapsto\Sigma'} \\
    \JGVExpr{\Gamma_2, x:t_1}{e_2}{t_2/\Sigma'\mapsto\Sigma''}
  }{
    \JGVExpr\Gamma{\Let{x}{e_1}{e_2}}{t_2/\Sigma\mapsto\Sigma''}
  }
}

\newcommand\ruleGVfork{
  \inferrule[T-Fork]{
    \JGVExpr\Gamma e \TUnit
  }{
    \JGVExpr\Gamma{\Fork e}\TUnit
  }
}
\newcommand\ruleGVforkEFF{
  \inferrule[T-Fork']{
    \JGVExpr[']\Gamma e {\TUnit/\Sigma\mapsto\emptyset}
  }{
    \JGVExpr[']\Gamma{\Fork e}{\TUnit/\Sigma\mapsto\emptyset}
  }
}
\newcommand\ruleGVsend{
  \inferrule[T-Send]{
    \JGVExpr{\Gamma_1}{e_1} t \\
    \JGVExpr{\Gamma_2}{e_2}{\Outp t s}
  }{
    \JGVExpr{\Gamma_1+\Gamma_2}{\Send {e_1}{e_2}} s
  }
}
\newcommand\ruleGVsendID{
  \inferrule[T-ID-Send]{
    \GVsplit\Gamma{\Gamma_1}{\Gamma_2} \\
    \JGVExprID{\Gamma_1}{e_1} t \\
    \JGVExprID{\Gamma_2}{e_2}{\Tagged\alpha{( \Outp t s)}}
  }{
    \JGVExprID\Gamma{\Send {e_1}{e_2}} {\Tagged\alpha s}
  }
}
\newcommand\ruleGVsendEFF{
  \inferrule[T-Send']{
    \GVsplit\Gamma{\Gamma_1}{\Gamma_2} \\
    \JGVExpr[']{\Gamma_1}{e_1} {d/\Sigma\mapsto\Sigma'} \\
    \JGVExpr[']{\Gamma_2}{e_2}{\Tagged\alpha{( \Outp d s)} / \Sigma'
      \mapsto \Sigma'',\alpha: \Outp d s }
  }{
    \JGVExpr[']\Gamma{\Send {e_1}{e_2}} {\Tagged\alpha s/
      \Sigma\mapsto \Sigma'',\alpha: s}
  }
}
\newcommand\ruleGVsendEFFS{
  \inferrule[T-Send'']{
    \GVsplit\Gamma{\Gamma_1}{\Gamma_2} \\
    \JGVExpr[']{\Gamma_1}{e_1} {\Tagged\beta{s'}/\Sigma\mapsto\Sigma'} \\
    \JGVExpr[']{\Gamma_2}{e_2}{\Tagged\alpha{( \Outp {s'} s)} / \Sigma'
      \mapsto \Sigma'',\alpha: \Outp t s, \beta:s' }
  }{
    \JGVExpr[']\Gamma{\Send {e_1}{e_2}} {\Tagged\alpha s/
      \Sigma\mapsto \Sigma'',\alpha: s}
  }
}
\newcommand\ruleGVrecv{
  \inferrule[T-Recv]{
    \JGVExpr\Gamma{e} {\Inp t s}
  }{
    \JGVExpr\Gamma{\Receive e} {t \lpair s}
  }
}
\newcommand\ruleGVrecvID{
  \inferrule[T-ID-Recv]{
    \JGVExprID\Gamma{e} {\Tagged\alpha{(\Inp t s)}}
  }{
    \JGVExprID\Gamma{\Receive e} {t \lpair {\Tagged\alpha s}}
  }
}
\newcommand\ruleGVrecvEFF{
  \inferrule[T-Recv']{
    \JGVExpr[']\Gamma{e} {\Tagged\alpha{(\Inp d s)} /
      \Sigma\mapsto\Sigma', \alpha:\Inp d s}
  }{
    \JGVExpr[']\Gamma{\Receive e} {d \lpair {\Tagged\alpha s} /
      \Sigma\mapsto \Sigma', \alpha: s}
  }
}
\newcommand\ruleGVrecvEFFS{
  \inferrule[T-Recv'']{
    \JGVExpr[']\Gamma{e} {\Tagged\alpha{(\Inp {s'} s)} /
      \Sigma\mapsto\Sigma', \alpha:\Inp {s'} s}
  }{
    \JGVExpr[']\Gamma{\Receive e} {{\Tagged\beta{s'}} \lpair {\Tagged\alpha s} /
      \Sigma\mapsto \Sigma', \alpha: s, \beta: s'}
  }
}
\newcommand\ruleGVaccept{
  \inferrule[T-Accept]{
    \JGVExpr\Gamma e {\Ap s}
  }{
    \JGVExpr\Gamma{\Accept e} s
  }
}
\newcommand\ruleGVacceptID{
  \inferrule[T-ID-Accept]{
    \JGVExprID\Gamma e {\Ap s} \\ \Fresh \alpha
  }{
    \JGVExprID\Gamma{\Accept e} {\Tagged\alpha s}
  }
}
\newcommand\ruleGVacceptEFF{
  \inferrule[T-Accept']{
    \JGVExpr[']\Gamma e {\Ap s / \Sigma \mapsto \Sigma'} \\ \Fresh\alpha
  }{
    \JGVExpr[']\Gamma{\Accept e} {\Tagged\alpha s / \Sigma \mapsto
      \Sigma', \alpha:s}
  }
}
\newcommand\ruleGVrequest{
  \inferrule[T-Request]{
    \JGVExpr\Gamma e{\Ap s}
  }{
    \JGVExpr\Gamma{\Request e}{\Dual s}
  }
}
\newcommand\ruleGVrequestID{
  \inferrule[T-ID-Request]{
    \JGVExprID\Gamma e{\Ap s} \\ \Fresh\alpha
  }{
    \JGVExprID\Gamma{\Request e}{\Tagged\alpha{\Dual s}}
  }
}
\newcommand\ruleGVrequestEFF{
  \inferrule[T-Request']{
    \JGVExpr[']\Gamma e{\Ap s / \Sigma \mapsto \Sigma'} \\ \Fresh\alpha
  }{
    \JGVExpr[']\Gamma{\Request e}{\Tagged\alpha{\Dual s} / \Sigma
      \mapsto \Sigma', \alpha:\Dual s}
  }
}
\newcommand\ruleGVclose{
  \inferrule[T-Close]{
    \JGVExpr\Gamma{e}{\End}
  }{
    \JGVExpr\Gamma{\Close e}{\TUnit}
  }
}
\newcommand\ruleGVcloseID{
  \inferrule[T-ID-Close]{
    \JGVExprID\Gamma{e}{\Tagged\alpha\End}
  }{
    \JGVExprID\Gamma{\Close e}{\TUnit}
  }
}
\newcommand\ruleGVcloseEFF{
  \inferrule[T-Close']{
    \JGVExpr[']\Gamma{e} {\Tagged\alpha\End /
      \Sigma\mapsto\Sigma', \alpha:\End}
  }{
    \JGVExpr[']\Gamma{\Close e} {\TUnit /
      \Sigma\mapsto \Sigma'}
  }
}
\newcommand\ruleGVnew{
  \inferrule[T-New]{\Unr\Gamma}{
    \JGVExpr\Gamma{\New s}{\Ap s}
  }
}
\newcommand\ruleGVnewEFF{
  \inferrule[T-New']{\Unr\Gamma}{
    \JGVExpr[']\Gamma{\New s}{\Ap s/\Sigma\mapsto\Sigma}
  }
}

\newcommand\ruleGVemp{
  \inferrule[T-Emp]{
    \Unr\Gamma
  }{
    \JGVExpr\Gamma \Remp {\Record{ }}
  }
}

\newcommand\ruleGVsingle{
  \inferrule[T-Single]{
    \JGVExpr\Gamma e t
  }{
    \JGVExpr\Gamma{\Rsingle\alpha e}{\Record{\alpha:t}}
  }
}

\newcommand\ruleGVconcat{
  \inferrule[T-Concat]{
    \JGVExpr{\Gamma_1}{e_1}{\{r_1\}} \\
    \JGVExpr{\Gamma_2}{e_2}{\{r_2\}} \\
    r_1 \sharp r_2
  }{
    \JGVExpr{\Gamma_1+\Gamma_2} {e_1 \cdot e_2}{\Record{r_1, r_2}}
  }
}

\newcommand\ruleGVsplitrecord{
  \inferrule[T-Field]{
    \JGVExpr\Gamma e{\Record{r, \alpha:t}}
  }{
    \JGVExpr\Gamma {\Rsplit e \alpha}{t \otimes \Record{r}}
  }
}

\newcommand\ruleGVsplitrecordmany{
  \inferrule[T-SplitRecord]{
    \JGVExpr\Gamma e{\Record{r_1 +r_2}} \\
    \Dom{r_1} = \alpha^*
  }{
    \JGVExpr\Gamma {\Rsplitmany e {\alpha^*}}{\Record{r_1} \otimes \Record{r_2}}
  }
}

\begin{document}
\title{Relating Functional and Imperative Session Types}
\author{Hannes Saffrich\lmcsorcid{0000-0002-1825-0097}}
\author{Peter Thiemann\lmcsorcid{0000-0002-9000-1239}}

\address{University of Freiburg, Germany}
\email{\{saffrich,thiemann\}@informatik.uni-freiburg.de}
%
% 150-250 words
\begin{abstract}
  Imperative session types provide an imperative interface to
  session-typed communication. In such an
  interface, channel references are first-class objects with
  operations that change the typestate of the channel.
  Compared to functional session type APIs, the program structure is simpler at the
  surface, but typestate is required to model the current state of
  communication throughout.

  Following an early work that explored the imperative approach, a
  significant body of work on session types has neglected the
  imperative approach and opts for a functional approach that uses
  linear types to manage channel references soundly. We
  demonstrate that the functional approach subsumes the early work on
  imperative session types by exhibiting a typing and semantics
  preserving translation into a system of linear functional session types.

  We further show that the untyped backwards translation from the
  functional to the imperative calculus is semantics
  preserving. We restrict the type system of the functional
  calculus such that the backwards translation becomes type
  preserving. Thus, we precisely capture the difference in expressiveness of the
  two calculi and conclude that the lack of expressiveness in the
  imperative calculus is largely due to restrictions imposed by its type system.
\end{abstract}
\maketitle              % typeset the header of the contribution

\section{Introduction}
\label{sec:introduction}

Session types provide a type discipline for  bidirectional
communication protocols in concurrent programs. They originate with papers by Honda and others~\cite{DBLP:conf/concur/Honda93,DBLP:conf/parle/TakeuchiHK94},
who proposed them as an expressive type system for binary
communication in pi-calculus. Later work considered embeddings in
functional and object-oriented languages, both theoretically and
practically oriented~\cite{DBLP:journals/jfp/GayV10,DBLP:conf/ecoop/HuKPYH10,DBLP:conf/ecoop/ScalasY16,DBLP:journals/jfp/Padovani17}.

A typical incarnation of session types~\cite{DBLP:journals/jfp/GayV10}
embedded in a functional language supports a data type of channel
ends, which are end points of communication channels.  A session type $s$ describes the communication behavior of a
channel end and is governed by a grammar like this one:
\begin{align*}
  s & ::= \Outp t s \mid \Inp t s \mid \oplus\{ \ell_i: s_i \} \mid
      \&\{ \ell_i : s_i \} \mid \End &
                                       t & ::= s \mid t \to t \mid
                                           t\otimes t \mid \dots
\end{align*}
Here, $t$ ranges over all types in the language (functions, pairs,
etc) including session types $s$. The session type $\Outp t s$
describes a channel on which we can send a value of type $t$ and then
continue communicating according to $s$. Dually, we can receive a
value of type $t$ and continue according to $s$ on a channel of type
$\Inp t s$. The internal choice type $\oplus\dots$ enables the sender
to choose the continuation protocol  $s_i$ by selecting  its label
$\ell_i$. The external choice
$\&\dots$ requires the receiver to continue with $s_i$ if it receives $\ell_i$. The session
type $\End$ marks the end of the conversation.

\subsubsection*{Functional vs Imperative Session Types}
\label{sec:funct-vs-imper}

A significant number of embedded session type systems rely on a
functional treatment of channel ends. That is, the communication
operations consume a channel end at the type before the communication
and conceptually produce a new channel end at the type after the
communication.
As an example we consider  the \KWReceive{} operation which consumes a channel of type
$\Inp t s$ and returns a pair of the
received value of type $t$ and the continuation channel of type $s$:
\begin{align*}
  \KWReceive &: \Inp t s \to (t \otimes s)
\end{align*}
This design forces a
programmer to explicitly thread channel references through the
program. Moreover, every channel reference must be treated linearly
because a repeated use at the same type would break the protocol.
The typical programming pattern is to rebind a variable, say
\lstinline/u/, containing the channel end with a different type in
every line as in \Cref{lst:example-functional} (typings refer to the state \emph{before} the
operation in that line).
\begin{lstlisting}[label={lst:example-functional},caption={Example
server in functional style},captionpos=b,float]
let server u =
  let (x, u) = receive u in    (* u: ?Int.?Int.!Int.s' *)
  let (y, u) = receive u in    (* u: ?Int.!Int.s' *)
  send (x+y, u)                (* u: !Int.s' *)
\end{lstlisting}
Writing a program in this style feels like functional programming
before the advent of monads, when programmers loudly complained
about the need for ``plumbing'' as demonstrated with \lstinline/u/. Moreover, this style is not
safe for embedding session types into mainstream languages because
most of them do not enforce the linearity needed to avoid aliasing
of channel ends at compile time. A similar argument can be made for
interfacing with dynamically typed languages.

There are techniques to ease the integration of linear functional APIs
in mainstream languages.
Embeddings in object-oriented languages make use of fluent interfaces,
which favor the chaining of method calls~\cite{DBLP:conf/fase/HuY16}.
Embeddings in functional languages wrap channels into a monad~\cite{DBLP:conf/haskell/PucellaT08}, but this approach either does not scale
well to programs that process multiple channels or it mimics an
imperative approach similar to what we explore in this paper. Generally, much less work
can be found that takes the alternative,
imperative approach inspired by typestate-based programming~\cite{DBLP:journals/tse/StromY86}.

Interestingly, in one of the early works on session types,
Vasconcelos, Gay, and Ravara~\cite{DBLP:journals/tcs/VasconcelosGR06} proposed
a session type calculus embedded in a multithreaded functional language, which we
call VGR\@.
It is a bit of a mystery why VGR was not called
imperative\footnote{The conference version of their paper~\cite{DBLP:conf/concur/VasconcelosRG04} 
is  called ``Session Types for Functional Multithreading''.}
because it enables rewriting the functional program fragment in
\Cref{lst:example-functional} into the style shown
\Cref{lst:example-server}, which clearly has an imperative flavor.
\begin{figure}[tp]
  \begin{minipage}[t]{0.49\linewidth}
\begin{lstlisting}[label={lst:example-server},caption={Example
server},captionpos=b]
fun server u =
  let x = receive u in
  let y = receive u in
  send x + y on u
\end{lstlisting}
  \end{minipage}
  \begin{minipage}[t]{0.49\linewidth}
\begin{lstlisting}[label={lst:example-server-unit},caption={Example server with capture},captionpos=b]
fun server' () =
  let x = receive u in
  let y = receive u in
  send x + y on u
\end{lstlisting}
  \end{minipage}
\end{figure}
The parameter \lstinline/u/ of the \lstinline/server/ function is a
reference to a communication channel.
The operation \lstinline/receive/ takes a channel associated with
session type $\Inp{\Int}S$ and returns an integer\footnote{Uppercase
  letters denote types in the VGR calculus.}. Executing
\lstinline/receive/ changes the type of the channel referred to by
\lstinline/u/ to $S$, which indicates that the VGR calculus is a typestate-based
system~\cite{DBLP:journals/tse/StromY86}. The function
\lstinline/send_on_/ takes an integer to transmit and a channel associated
with session type $\Outp{\Int}S$. It returns a unit value and updates
the channel's type to $S$.

Taken together, the \lstinline/server/ function in \Cref{lst:example-server} expects
that its argument \lstinline/u/ refers to a channel of type
$\Inp\Int\Inp\Int\Outp\Int S'$, for some $S'$, and leaves it in a state corresponding
to type $S'$ on exit. This
change of typestate is reflected in the shape of a function type in VGR\@:
$
  \Sigma_1; T_1 \to T_2; \Sigma_2
$. In this type, $T_1$ and $T_2$ are argument and return type of the
function. The additional components $\Sigma_1$ and $\Sigma_2$ are environments that reflect
the state (session type) of the channels before ($\Sigma_1$) and after ($\Sigma_2$)
calling the function. The type of a channel, $\Chan\Nchan$, serves as a pointer to
the entry for $\Nchan$ in the current channel environment $\Sigma$. Channels in $T_1$
refer to entries in $\Sigma_1$ and channels in $T_2$ refer to entries
in $\Sigma_2$, but both environments may refer to further channels
that describe channel references captured by the function ($\Sigma_1$)
or created by the function ($\Sigma_2$). In \Cref{lst:example-server}, the
type of \lstinline/server/ is
\begin{gather}\label{eq:1}
  \{ \Nchan : \Inp\Int\Inp\Int\Outp\Int S \} ; \Chan\Nchan \to \TUnit;
  \{ \Nchan : S \} \text,
\end{gather}
for some fixed channel name $\Nchan$ and session type $S$.

Compared to other session type systems~\cite{DBLP:journals/jfp/GayV10,DBLP:journals/pacmpl/FowlerLMD19}, VGR
does \textbf{not} require linear handling of channel references, as
can be seen by the multiple uses of variable \lstinline/u/ in
\Cref{lst:example-server}. Instead, it keeps track of the
current state of every channel using the environment $\Sigma$, which
is threaded linearly through the typing rules.

In \Cref{sec:motivation} we give deeper insights into VGR, the kind of
programs that it accepts, and the programs that fail to typecheck.
To give a glimpse of its peculiarities, we examine the type of
\lstinline/server/ in~\cref{eq:1} more closely.

First, the type $\Chan\Nchan$ of a channel reference refers to the \emph{name} $\Nchan$. This
name identifies a certain
channel so that the function cannot be invoked on other channels.
Second, a function of this type can be type-checked without knowledge of
the channel names that are currently in use and their state. This property enables the
definition of the \lstinline/server/ function in a library, say, but
the type checker does not allow us to call the function on a channel
named differently than $\Nchan$, even if its session type
matches. Hence, the library may end up defining a function that cannot be called.

Consider the variation of the type in~\cref{eq:1} that replaces the argument
type by $\TUnit$:
\begin{gather}\label{eq:2}
  \{ \Nchan : \Inp\Int\Inp\Int\Outp\Int S \} ; \TUnit \to \TUnit;
  \{ \Nchan : S \} \text.
\end{gather}
This type can be assigned to a function like \lstinline/server'/ in
\Cref{lst:example-server-unit} that is closed over a
reference to a channel of type $\Chan\Nchan$. In this context, the
fixation on a certain channel name $\Nchan$ is required for
soundness: While we might want to apply a function to different
channels, it is not possible to replace a channel captured in a
closure. A function of type as in \cref{eq:2} may be called any time the channel
$\Nchan$ is in a state matching the ``before'' session type of the
function.

Subsequently, Gay and Vasconcelos
created a functional session type
calculus based on a linear type system, which was later called LFST\footnote{Linear
  Functional Session Types.}~\cite{DBLP:journals/jfp/GayV10}. While LFST is still monomorphic, a
function like \lstinline/server/ can be applied to several different
channels with the same session type.
In LFST, we can also close over a channel, but doing so turns a
function like \lstinline/server'/ into a function that \textbf{must be called
exactly once}. In
contrast, \lstinline/server'/ can be called arbitrarily often
(including zero times) in VGR provided
the channel $\Nchan$ is
available at the right type in the caller's environment.
Clearly, LFST lifts some restrictions of the VGR calculus,
but it seems to impose other restrictions. In any case, the exact
correspondence between the two calculi has never been studied.

There is another line of session-type research based on the
Curry-Howard correspondence between fragments of linear logic and
process calculi~\cite{DBLP:conf/concur/CairesP10}. Programs/processes
in these systems may also be regarded as handling channels
``imperatively'', perhaps even more so than VGR\@. We discuss these
approaches in \Cref{sec:related-work} along with other related
work.

\subsection*{Contributions}
\label{sec:contributions}

\begin{itemize}
\item We show  that LFST is at least as expressive
  as VGR by giving a  typing-preserving translation that simulates
  VGR in LFST (\Cref{sec:transl-imper-funct}).
\item We show that untyped VGR is at least as expressive as LFST by
  giving a backwards translation that simulates LFST in VGR (\Cref{sec:backw-transl}).
\item We exhibit a type system for LFST that characterizes the
  shortcomings of VGR exactly. The backwards translation becomes type
  preserving with respect to this system (\Cref{sec:towards-typed-backw}).
\end{itemize}

In this paper we omit choice and recursion from session types because these features
are straightforward to add and our results extend seamlessly.
Compared to the conference version of this paper~\cite{DBLP:conf/coordination/Saffrich021}, we added more 
explanations, we incorporated full rule sets and proofs, and we made
the Agda proof script for
\Cref{thm:translation-preserves-typing} (translation
preserves typing) available as a supplement~\cite{saffrich_hannes_2021_5764782}.

\section{Motivation}
\label{sec:motivation}

In this section, we highlight the various shortcomings of VGR and
discuss how they are solved in LFST\@.

\subsection{Channel Identities}
\label{sec:channel-abstraction}

Our discussion of VGR's function type $\Sigma_1;T_1 \to
T_2;\Sigma_2$ in the introduction shows that a function that takes a
channel as a parameter can only be applied to a single channel. A
function like \lstinline/server/ (\Cref{lst:example-server}) must be applied to the channel of
type $\Chan\Nchan$, for some fixed name $\Nchan$.

LFST sidesteps this restriction by not encoding the identity of a channel in
the type. It rather posits that session types are linear so that
channel references cannot be duplicated. In consequence, the
operations of LFST's session API must consume a channel and return another channel to continue the protocol.

\subsection{Data Transmission vs Channel Transmission}
\label{sec:data-transmission-vs}

In VGR, it is possible to pass channels
from one thread to another. The session type $\Outp{S'}{S}$ indicates a higher-order
channel on which we can send a channel of type $S'$. The operation to send
a channel has the following typing rule in VGR\@:
\begin{mathpar}
  \ruleVGRSendS
\end{mathpar}
The premises are \emph{value typings} that indicate that $v$ and $v'$ are
references to different, fixed channels $\beta$ and $\alpha$ under variable environment
$\Gamma$. The conclusion is an
\emph{expression typing} of the form $\JVGRExpr\Gamma\Sigma
e{\Sigma_1}{T}{\Sigma_2}$ where $\Sigma$ is the incoming channel
environment, $\Sigma_1$ is the part of $\Sigma$ that is passed
through without change, and $\Sigma_2$ is the outgoing channel environment after the
operation indicated by expression $e$ which returns a result of type $T$.
The rule states that channels $\alpha$ and $\beta$ have session type
$\Outp{S'}{S}$ and $S'$, respectively. Moreover, $\alpha$ and $\beta$
are implicitly different (and different from all names in $\Sigma$)
because a channel environment is only
well-formed if all its bound names are different. The channel $\beta$ is consumed
(it is sent to the other end of channel $\alpha$) and $\alpha$
gets updated to session type $S$.

Compared to the function type considered in the introduction (\Cref{sec:introduction}), sending a channel is more flexible.
Any channel of type $S'$ can be passed because $\beta$ is not part
of channel $\alpha$'s session type. Alas, if the sender  holds
references to channel $\beta$ (i.e., values of type $\Chan\beta$),
then these references can no longer be
exercised as $\beta$ has been removed from $\Sigma$.
So one can say that rule \TirName{C-SendS} passes ownership of channel
$\beta$ from the sender to the receiver.

However, there is another way to send a channel reference over a
channel, namely if it is captured in a closure. To see what happens in
this case, we look at VGR's typing rules for sending and receiving
data of type $D$. Types of the form $D$ comprise first-order types and
function types, but not channels.
\begin{gather*}
  \ruleVGRSendD
  \quad
  \ruleVGRReceiveD
\end{gather*}
One possibility for type $D$ is a function type like
$D_1 = \{\beta: S'\}; \TUnit \to \TUnit; \{\beta:S''\}$. A function of this type
captures a channel named $\beta$ which may or may not occur in
$\Sigma$. It is instructive to see what happens at the receiving end
in rule \TirName{C-ReceiveD}.
If we receive a function of type $D_1$ and $\Sigma$ already contains
channel $\beta$, then we will be able to
invoke the function as soon as the type of $\beta$ is $S'$ (if ever).

If channel $\beta$ is not yet present at the receiver, it turns out we
cannot send it in a subsequent communication, as the received channel
gets assigned a fresh name $d$:
\begin{mathpar}
  \ruleVGRReceiveS
\end{mathpar}
For the same reason, it is impossible to send channel $\beta$ first
and then the closure that refers to it: the name $\beta$ gets removed
from the sender's channel environment, the receiver renames $\beta$ to some
fresh $d$, and the closure that remains with the sender still refers to $\beta$.
At this point, the sender cannot invoke the closure anymore as $\beta$
is required but does not exists in the sender's channel environment.
Thus, sending the channel first effectively cuts all previous connections.

We conclude with the observation that abstracting over the \KWSend{}
operation is not usefully possible in VGR because it would fix channel
names in the function type.

None of these issues arise in LFST because channels have no
identity. Hence, any value whatsoever can be sent over a channel,
higher-order session types are possible, and there is just one typing
rule for sending and another for receiving any kind of value.

\subsection{Channel Aliasing}
\label{sec:channel-aliasing}

The VGR paper discusses the following function \lstinline/sendSend/.
%[label={lst:example-aliasing},caption={Channel aliasing},captionpos=b]
\begin{lstlisting}
fun sendSend u v = send 1 on u; send 2 on v
\end{lstlisting}
It takes two channels and sends a
number on each. This use is reflected in the following typing.
\begin{gather}\label{eq:3}
  \mathtt{sendSend} :
 \Sigma_1; \Chan u \to (\Sigma_1; \Chan v \to \TUnit; \Sigma_2); \Sigma_1
\end{gather}
with $\Sigma_1 = \{u : \Outp\Int{S_u}, v : \Outp\Int{S_v}\}$
and  $\Sigma_2 = \{u : {S_u}, v : {S_v}\}$.

Ignoring the types we observe that it would be semantically sound to pass a reference to
the same channel \lstinline/w/, say, of session type \lstinline/!Int.!Int.End/
for \lstinline/u/ and \lstinline/v/. However, \lstinline/sendSend w w/
does not type check with the type in~\cref{eq:3} because \lstinline/w/ would
have to have identity $u$ and $v$ at the same time, but  environment
formation mandates they must be different.

Another typing of \lstinline/sendSend/ in VGR would be
\begin{gather}
  \label{eq:30}
  \mathtt{sendSend}' :
  \Sigma_1;\Chan w \to (\Sigma_1; \Chan w \to \TUnit; \Sigma_2);
  \Sigma_1
\end{gather}
with $\Sigma_1 = \{ w : \Outp\Int{\Outp\Int{S_w}} \}$ and
$\Sigma_2 = \{ w : S_w \}$. With this typing, \lstinline/sendSend w w/
type checks. Indeed, the typing forces the two arguments to be aliases!

In LFST, the invocation \lstinline/sendSend w w/ is not legal as it
violates linearity. Indeed, to simulate the two differently typed
flavors of \lstinline/sendSend/ requires two different expressions in
LFST\@. As an illustration, we show LFST expressions as they are
produced by our type-driven translation in
\Cref{sec:transl-imper-funct}, when applied to the
\lstinline/sendSend/ function with the types in~\eqref{eq:3} and
in~\eqref{eq:30}.
\begin{figure}[tp]
  \begin{minipage}{0.49\linewidth}
\begin{lstlisting}[label={lst:example-trans-noalias},caption={Without aliasing},captionpos=b,basicstyle=\scriptsize\ttfamily]
fun sendSend u v sigma =
  let (cu, sigma) = sigma.u in
  let cu' = send 1 on cu in
  let sigma = sigma * {u: cu'} in
  let (cv, sigma) = sigma.v in
  let cv' = send 2 on cv in
  let sigma = sigma * {v: cv'} in
  ((), sigma)
\end{lstlisting}
  \end{minipage}
  \begin{minipage}{0.49\linewidth}
\begin{lstlisting}[label={lst:example-trans-alias},caption={With aliasing},captionpos=b,basicstyle=\scriptsize\ttfamily]
fun sendSend' u v sigma =
  let (cw, sigma) = sigma.w in
  let cw' = send 1 on cw in
  let sigma = sigma * {w: cw'} in
  let (cw, sigma) = sigma.w in
  let cw' = send 2 on cw in
  let sigma = sigma * {w: cw'} in
  ((), sigma)
\end{lstlisting}
  \end{minipage}
%  \vspace{-1.5\baselineskip}
%  \caption{Translated code fragments}
  \label{fig:translated-code-fragments}
\end{figure}
In the code fragment in \Cref{lst:example-trans-noalias}, \lstinline/u/ and \lstinline/v/ have unit type
(translated from $\Chan u$ and $\Chan v$) and \lstinline/sigma/ is a
linear record with fields \lstinline/u/ and \lstinline/v/ that contain
the respective channels. The dot operator performs field selection and
\lstinline/*/ is disjoint record concatenation. The notation for
record literals is standard.

In the translation of \lstinline/sendSend'/ in
\Cref{lst:example-trans-alias}, \lstinline/u/ and \lstinline/v/
also have unit type (translated from $\Chan w$ and $\Chan w$), but the
record
\lstinline/sigma/ has only one field \lstinline/w/ containing the channel.

\subsection{Abstraction over Channel Creation}
\label{sec:abstr-over-chann}

A server typically accepts many connections on the same access point
and performs the same initialization (e.g., authentication) on each
channel. Hence, it makes sense to abstract over the creation of a
channel as in this code fragment.
\begin{lstlisting}
fun acceptAdd () =
  let c = accept addService in
  // authenticate client on c (omitted)
  c
\end{lstlisting}
Here, \lstinline/addService/ is an access point for  sessions of type $S$. The
function \lstinline/accept/  creates a channel end according to the
access point type (a client would invoke the corresponding \lstinline/request/
function on the same access point.) The VGR typing rule for accepting a
connection reads as follows:
\begin{gather*}
  \ruleVGRAccept
\end{gather*}
In this rule, $v$ is an access point for creating connections of type $S$.
According to the rule, the name of the newly created channel is fresh,
i.e., it does not occur in any incoming environment or type.
However, the freshness condition on this channel only applies inside
the function body of \lstinline/acceptAdd/. The actual
VGR type of \lstinline/acceptAdd/ does not reflect freshness anymore
but fixes a name $\alpha$, say, in the function type:
\begin{gather*}
  \{\}; \TUnit \to \Chan\alpha; \{\alpha: \Inp\Int{\Inp\Int{\Outp\Int{S'}}} \}
\end{gather*}
In consequence, VGR cannot invoke \lstinline/acceptAdd/ twice in a row
as the second invocation would result in an ill-formed environment that
contains two specifications for channel $\alpha$.

LFST elides this issue, again, by not tracking channel identities.

\section{Two Session Calculi}
\label{sec:two-session-calculi}

This section formally introduces the calculi VGR and LFST\@. It also
explains the slight adjustments to the calculi that we made to obtain
a smooth translation.

\subsection{VGR\@: Imperative Session Types}
\label{sec:vgr}

\begin{figure}[tp]
  \begin{minipage}{0.49\linewidth}
  \begin{align*}
    C &::=\Thread t \mid (C \| C) \mid (\nu x:\Ap S)C \mid
        (\nu\gamma)C \\
    t &::= v \mid \Let xet \mid \Fork t;t \\
    e &::= t \mid v\,v \mid \New S \mid \Accept v \mid \Request v\\
    & \mid
        \Send vv \mid \Receive v \mid \Close v \\
    v &::= \alpha \mid \Lam {(\Sigma; x:T)} e \mid % \Rec{x:T} v \mid
        \UnitV \\
    \alpha &::=  x \mid \gamma^p \\
    p &::= + \mid - 
  \end{align*}
  \end{minipage}
  \begin{minipage}{0.49\linewidth}
    \begin{align*}
    T &::= D \mid \Chan \alpha \\
    D &::= \Ap S \mid \Sigma;T \to T; \Sigma \mid \TUnit \\
    S &::= \Inp D S \mid \Outp D S \mid \Inp S S \mid \Outp S S
        \mid \End \\
    \Sigma &::= \emptyset \mid \Sigma, \alpha:S \qquad
             (\alpha\notin\Sigma) \\
    \Gamma &::= \emptyset \mid \Gamma, x : T \qquad (x \notin \Gamma)
    \end{align*}
  \end{minipage}
  \caption{Syntax of VGR}
  \label{fig:syntax-vgr}
\end{figure}

%%% Local Variables:
%%% mode: latex
%%% TeX-master: "main"
%%% End:

\begin{figure}[tp]
  Evaluation contexts
  \begin{align*}
    E & ::= \Hole \mid \Let x e t
  \end{align*}
  Reduction of expressions and processes \quad
    \fbox{$t \VGREReduceTo[\RLabel] t$} \quad \fbox{$C
    \VGRPReduceTo[\RLabel] C$}
  \begin{gather}
    \label{eq:25}
    E[(\Lam{(\Sigma;y:T)} e)v]
    \VGREReduceTo
    E[e[v/y]]
    \\
    % \label{eq:26}
    % E[\Let y e t]
    % \VGREReduceTo
    % \Let y e {E[t]}
    % \\
    \label{eq:27}
    E[\Let x v t]
    \VGREReduceTo
    E[t[v/x]]
    \\
    \label{eq:29}
    {E[{\Request n}]}
    \VGREReduceTo[\Request \gamma]
    {E[{\gamma^+}]}
    \qquad
    {E[ {\Accept n}]}
    \VGREReduceTo[\Accept\gamma]
    {E[{\gamma^-}]}
    \\
    \label{eq:31}
    {E[{\Receive {\gamma^p}}]}
    \VGREReduceTo[\gamma^p ? v]
    {E[v]}
    \qquad
    {E[{\Send v {\gamma^p}}]}
    \VGREReduceTo[\gamma^p ! v]
    {E[\UnitV]}
    \\
    {E[{\Close {\gamma^p}}]}
    \VGREReduceTo[\gamma^p \End]
    {E[\UnitV]}
    \\
    \label{eq:21}
    \inferrule{
      t_1 \VGREReduceTo[\KWRequest \gamma] t_1' \\
      t_2 \VGREReduceTo[\KWAccept \gamma] t_2'
    }{
      \Thread{t_1} \| \Thread{t_2}
      \VGRPReduceTo[\KWAccept]
      \Cnewap\gamma
      \Thread{t_1'} \| \Thread{t_2'}
    }
    \qquad
    \inferrule{
      t_1 \VGREReduceTo[\gamma^p ? v] t_1' \\
      t_2 \VGREReduceTo[\gamma^{\Dual p} ! v] t_2'
    }{
      \Thread{t_1} \| \Thread{t_2}
      \VGRPReduceTo[\KWSend]
      \Thread{t_1'} \| \Thread{t_2'}
    }
    \\
    \inferrule{
      t_1 \VGREReduceTo[\gamma^p \End] t_1' \\
      t_2 \VGREReduceTo[\gamma^{\Dual p} \End] t_2'
    }{
      \Thread{t_1} \| \Thread{t_2}
      \VGRPReduceTo[\KWClose]
      \Thread{t_1'} \| \Thread{t_2'}
    }
    \\
    \label{eq:23}
    \Thread{E[\New\,S]}
    \VGRPReduceTo[\KWNew]
    \Cnewap{ n: \Ap S} \Thread{E[n]}
    \\
    \label{eq:24}
    \Thread{E[\Fork t_1;t_2]}
    \VGRPReduceTo[\KWFork]
    \Thread{t_1} \| \Thread{E[t_2]}
    \\
    \label{eq:28}
    \inferrule{t \VGREReduceTo t'}{\Thread t \VGRPReduceTo
      \Thread{t'}}
    \quad
    \inferrule{C \VGRPReduceTo[\RLabel] C'}{\Cnewap\gamma C \VGRPReduceTo[\RLabel]
      \Cnewap\gamma C'}
    \quad
    \inferrule{C \VGRPReduceTo[\RLabel] C'}{\Cnewap{n:T} C \VGRPReduceTo[\RLabel]
      \Cnewap{n:T} C'}
    \quad
    \inferrule{C \VGRPReduceTo[\RLabel] C'}{
      C\|C'' \VGRPReduceTo[\RLabel]
      C'\|C''}
  \end{gather}
  \caption{Semantics of VGR}
  \label{fig:semantics-vgr}
\end{figure}

%%% Local Variables:
%%% mode: latex
%%% TeX-master: "main"
%%% End:

\begin{figure}[tp]
  \begin{mathpar}
    \ruleVGRConst

    \ruleVGRChan

    \ruleVGRVar

    \ruleVGRAbs
    %
    % \ruleVGRRec
  \end{mathpar}
  \caption{Value typing rules of VGR \fbox{$\JVGRValue\Gamma v T$}}
  \label{fig:value-typing-vgr}
\end{figure}

%%% Local Variables:
%%% mode: latex
%%% TeX-master: "main"
%%% End:

\begin{figure}[tp]
  \begin{mathpar}
    \ruleVGRReceiveD

    \ruleVGRReceiveS

    \ruleVGRSendD

    \ruleVGRSendS

    \ruleVGRClose

    \ruleVGRAccept

    \ruleVGRRequest

    \ruleVGRVal

    \ruleVGRApp

    \ruleVGRNew

    \ruleVGRLet

    \ruleVGRForkRevised
  \end{mathpar}
  \caption{Expression typing rules of VGR \fbox{$\JVGRExpr\Gamma\Sigma
      e {\Sigma'}T{\Sigma''}$}}
  \label{fig:expression-typing-vgr-full}
\end{figure}

%%% Local Variables:
%%% mode: latex
%%% TeX-master: "main"
%%% End:
%{fig-vgr-expression-typing-excerpt}
\begin{figure}[tp]
  \begin{mathpar}
    \ruleVGRThread

    \ruleVGRPar

    \ruleVGRNewN

    \ruleVGRNewB

    \ruleVGRNewC
  \end{mathpar}
  \caption{Configuration typing rules of VGR}
  \label{fig:config-typing-rules-vgr}
\end{figure}

%%% Local Variables:
%%% mode: latex
%%% TeX-master: "main"
%%% End:

\Cref{fig:syntax-vgr} defines the syntax of VGR~\cite{DBLP:journals/tcs/VasconcelosGR06}. 
Processes $C$ are expression
processes, parallel processes, protocol restrictions, and channel
restrictions, in that order. Expressions $t$ are in
A-normal form, i.e., they are
sequences of simple expressions $e$ ending in a fork that creates new
threads or in a value. A simple expression $e$ restricts all its arguments to
values, complex expressions must be sequentialized by using
let-expressions. Simple expressions are function application, access point creation,
accepting and requesting a connection, sending and receiving on a
channel, and closing a channel.  A value $v$ is either a channel name $\alpha$, a
lambda abstraction, or a unit value. Channel names are either
variables $x$ or channel ends $\gamma$ with a polarity $p$.
Types distinguish between data types $D$ and channels
because two different sets of typing rules govern sending and
receiving of data vs. sending and receiving a channel. We already used this
syntax informally in the examples. It is folklore that any expression can be transformed into
A-normal form (see also \Cref{sec:untyped-translation}).

We write $\Dual{\parbox[c]{0.5em}{$\,\cdot\,$}}$
for the \emph{dual operator}. It flips the polarity of
a communication. On polarities, it is defined as $\Dual{+} = -$ and
$\Dual{-} = +$. On session types, $\Dual{\Inp TS} = \Outp T\Dual S$
and $\Dual{\Outp TS} = \Inp T\Dual S$. In both cases the dual
operator is an involution: $\Dual{\Dual S} = S$.

We omit choices as they
present no significant problem and as they can be simulated using
channel passing. We also omit the standard congruence rules for
processes and silently apply reduction rules up to congruence:
parallel composition is a commutative monoid, the
$\nu$-binders admit scope extrusion, and $\nu$-binders can commute.

\Cref{fig:semantics-vgr} defines the semantics of VGR\@.  We use a
slightly different, but equivalent definition as in the literature. We define evaluation
contexts for expressions $E, F ::= \Hole \mid \Let x E t$ which are used in
the expression rules. Our formulation avoids the commuting conversion rule \TirName{R-Let} in the
literature and fixes an issue with the original reduction
relation.\footnote{$\Let x {\Fork t; t'} t''$ is stuck in the original
work~\cite{DBLP:journals/tcs/VasconcelosGR06}.} We distinguish between expression reduction
$\VGREReduceTo[\RLabel]$ and process reduction
$\VGRPReduceTo[\RLabel]$, both of which are tagged with a label
$\RLabel$. This label indicates the effect of
the reduction and it ranges over
\begin{align*}
  \RLabel &::= \KWAccept \mid \KWSend \mid \KWClose \mid \KWNew \mid \KWFork \mid
            \ELabel & \text{processes} \\
  \RLabel &::= \Accept{\gamma^p} \mid \Request{\gamma^p} \mid \gamma^p?v
            \mid \gamma^p!v \mid \gamma^p\End \mid \ELabel & \text{expressions}
\end{align*}
where $\ELabel$ stands for effect freedom and can be omitted. Labeled
expression reductions are paired with their counterpart at the process
level as familiar from process calculi~\cite{DBLP:books/daglib/0098267}, that is, $\gamma^p?v$ ($\gamma^p!v$) stand for
receiving (sending) $v$ on $\gamma^p$ which resolves to label
$\KWSend$ at the process level (see reduction~\eqref{eq:21}). Similarly,
$\Accept{\gamma^p}$ ($\Request{\gamma^p}$) stands for accepting (requesting) a
connection on fresh channel $\gamma^p$ and resolves to label $\KWAccept$
at the process level. Finally, $\gamma^p\End$ stands for a close
operation on $\gamma^p$ and resolves to label $\KWClose$ at the process level.

Typing for VGR comes in three parts: value typing
$\JVGRValue\Gamma v T$ in \Cref{fig:value-typing-vgr}, expression
typing $\JVGRExpr\Gamma\Sigma e {\Sigma'} T{\Sigma''}$ in
\Cref{fig:expression-typing-vgr-full}, and configuration typing
$\JVGRConfig\Gamma\Sigma C{\Sigma'}$ (\Cref{fig:config-typing-rules-vgr}).
The value typing judgment relates an environment $\Gamma$ and a value
$v$ to a type $T$. The expression typing judgment is very similar to a
type state system. It relates a typing environment $\Gamma$, an
incoming channel environment $\Sigma$, and an expression to an
environment $\Sigma' \subseteq \Sigma$ which contains the channels not
used by $e$, the type $T$, and the outgoing channel environment
$\Sigma''$. $\Sigma''$ contains typings for channels that have been
used by $e$ or created by $e$. The configuration typing relates
$\Gamma$, incoming $\Sigma$, and configuration $C$ with $\Sigma'
\subseteq \Sigma$ which contains the channels not used by $C$.

The static semantics of the VGR calculus is presented as in
the literature~\cite{DBLP:journals/tcs/VasconcelosGR06} except for the
rule \TirName{C-Fork}. This change is unavoidable because the original
rule is unsuitable for the translation:
\begin{mathpar}
  \ruleVGRFork
\end{mathpar}
It states that $t_1$ processes channels in $\Sigma$, leaves the
channels in $\Sigma_1 \subseteq \Sigma$ unchanged, and consumes the
remaining ones. The unchanged channels $\Sigma_1$ are then processed
by $t_2$. However, the translation of $t_1$ runs in a separate thread,
so it is  unable to return the untouched channels in
$\Sigma_1$. Hence, the rule \TirName{C-Fork} splits the channels into
the ones in $\Sigma_1$ consumed by the new thread $t_1$ and the ones
in $\Sigma_2$ consumed by the continuation $t_2$. 

We also deviate in
using a labeled transition system for the dynamic semantics to
directly relate labeled reduction steps between the two systems.

%% PJT: explain some typing rules?

\subsection{Linear Functional Session Types}
\label{sec:line-funct-sess}
\begin{figure}[t]
  \begin{align*}
    & \mathrm{Constants}&
                         k &::= \Fix \mid \Fork \mid \SendF \mid \Receive \mid \Accept \mid
        \Request \mid \Newinline{$\Close \mid \New$} \\
    & \mathrm{Expressions}&
    e &::= x \mid \alpha \mid k \mid \UnitV \mid \Lam x e \mid e~e \mid (e,
        e) \mid \Let{(x,y)}ee\\
                      &&&\mid  \Newinline{$ \Let{()}ee \mid \Remp \mid \Rsingle \alpha e \mid
        \Rconcat e e \mid \Rsplit e \alpha \mid \Rsplitmany e
        {\vec\alpha}$} \\
    & \mathrm{Configurations}&
    C &::= \Thread e \mid C\|C \mid \Cnewchan\gamma\delta C \mid
        \Newinline{$\Cnewap n C$}\\
    & \mathrm{Types}&
    t &::=  s \mid \Ap s \mid \TUnit \mid t \to t \mid t \linto t \mid t\lpair t \mid \Newinline{$\Record{r}$} \\
    & \mathrm{Session Types}&
    s &::= \Inp t s \mid \Outp t s \mid \End \\
    & \mathrm{Rows}&
    r &::= \Rownull \mid r, \alpha : t \\
    & \mathrm{Environments}&
    \Gamma &:= \TEnull \mid \Gamma, \alpha : t \mid \Gamma, x : t
  \end{align*}
  \caption{Syntax of LFST}
  \label{fig:lfst-syntax}
\end{figure}
\begin{figure}[tp]
  \input{mathpar-gv-unrestricted-types}
  \hrule
  \input{mathpar-gv-splitting}
  \hrule
  \begin{mathpar}
    \ruleGVvar

    \ruleGVlamunr

    \ruleGVlamlin

    \ruleGVapp

    \ruleGVappLin
    \\
    \ruleGVunit

    \ruleGVunitE

    \ruleGVpair

    \ruleGVpairE
  \end{mathpar}
  \input{mathpar-gv-record-typing-rules}
  \caption{Typing rules of LFST}
  \label{fig:lfst-typing-record-operations}
\end{figure}
On the functional side, we consider an extension of a
synchronous variant of the LFST calculus~\cite{DBLP:journals/jfp/GayV10} by
linear records with disjoint
concatenation. \Cref{fig:lfst-syntax} gives the syntax of this
calculus, which we call LFST-rec.
The syntax is taken from the literature, except for the cases with gray
background color, which were added to match the VGR calculus.
The $\New$-constant creates an access point, which is bound by a $\Cnewap n C$ configuration.
The $\Close$-constant closes a channel of session type $\End$.
The second line of the expression
grammar adds the standard elimination of linear units and defines operations on linear records. We write
$\Remp$ for the empty record, $\Rsingle\alpha e$ to construct a singleton record with field
$\alpha$ given by $e$, $\Rconcat {e_1}{e_2}$ for the disjoint
concatenation of records $e_1$ and $e_2$, $\Rsplit e \alpha$ to project
field $\alpha$ out of the record $e$  returning a pair of the
contents of the field and the remaining record, and $\Rsplitmany e
{\vec\alpha}$ for generalized projection to a list of names $\vec\alpha$ that
returns a pair of two records, one with the fields $\vec\alpha$ and the
other with the remaining fields.

The extension with records can be regarded as syntactic
sugar as it is well known how to compile records to nested
pairs. Given that compilation, the typing rules for record operations
are derived rules. We
prefer the convenience of the record notation as it avoids the
additional bookkeeping of this compilation step.

A configuration $C$ can be a single thread, two configurations running
in parallel, a channel abstraction binding the two ends to $\gamma$
and $\delta$, or an access point abstraction $\Cnewap n C$. The latter is a
straightforward addition to LFST, which assumes the existence of
globally known access points.

The metavariable $t$ ranges over types, $s$ ranges over session types, and $r$ ranges over rows, which are lists of
bindings of names to types. A type $t$ can be a session type $s$, a
access point type $\Ap s$,
the unit type $\TUnit$, an unrestricted function type $t \to t$, a single-use function
type $ t \linto t$, a pair type $t \lpair t$, or a record type
$\Record r$ defined by a row $r$. A session
type $s$ is as before. A row $r$ is a list of pairs of (row) names and
types where all names are disjoint.

\Cref{fig:lfst-typing-record-operations} recalls the definition
of the predicate $\Unr t$ for unrestricted types, which we lift pointwise to typing
environments.  Intuitively, a type is unrestricted if it does not
contain any linear components. A linear component is either a session
type; a pair with at least one linear component; a record with at
least one linear field; or a single-use function, which may close over a
linear component in a free variable.
In the literature~\cite{DBLP:journals/jfp/GayV10}, a 
channel of type $\End$ is unrestricted, so that no explicit $\Close$
operation is needed. Here, we fully enforce linear handling of
channels by adding the $\Close$ operation. This addition requires a change in the  $\Unr t$
predicate.

\Cref{fig:lfst-typing-record-operations} also recalls the
splitting judgment $\GVsplit\Gamma{\Gamma_1}{\Gamma_2}$. It splits 
environment $\Gamma$ into $\Gamma_1$ and $\Gamma_2$ such that
unrestricted bindings are duplicated and linear bindings end up either
in $\Gamma_1$ or $\Gamma_2$. In the typing rules, we write
$\Gamma_1+\Gamma_2$ for some $\Gamma$ such that
$\GVsplit\Gamma{\Gamma_1}{\Gamma_2}$. 

\Cref{fig:lfst-typing-record-operations} also contains the well-known typing rules for the communication
primitives as well as the (derived) rules for the record fragment of LFST\@. The rule \TirName{T-Emp}
typechecks the empty record with the premise $\Unr\Gamma$ which states
that $\Gamma$ only contains unrestricted types. The rule
\TirName{T-Single} is unsurprising. Premise $
\GVsplit\Gamma{\Gamma_1}{\Gamma_2}$ of rule \TirName{T-Concat} splits
the incoming environment $\Gamma$ so that bindings to a linear type end
up either in $\Gamma_1$ or in $\Gamma_2$ (also in
\Cref{fig:lfst-typing-record-operations}). Premise $r_1 \sharp
r_2$ states that rows $r_1$ and $r_2$ are disjoint, which means they bind
different field names. Under these assumptions the (disjoint) concatenation of records
$e_1$ and $e_2$ is accepted.

The rules for field access and splitting of the record generalize the
elimination rule for linear pairs. Rule \TirName{T-Field} shows
that  a field access singles out the field named $\alpha$. Its content
is paired up with a record comprising the remaining fields. Linearity
of the record's content is preserved as the pair is also linear. Rule
\TirName{T-SplitRecord} is similar, but splits its subject $e$ according
to a list $\vec\alpha$ of names which must be present in $e$'s type. The result
is a linear pair of two records. We consider an empty record to be
unrestricted so that we can drop it if needed.

The remaining typing rules are taken from the original paper~\cite{DBLP:journals/jfp/GayV10}. 
We modify the operational semantics
to perform synchronous communication and
to fit with the labeled transition style used for VGR in
\Cref{sec:vgr}. Its formalization is omitted from the main text because of its
similarity to VGR, but it is available in the appendix
(\Cref{fig:semantics-lfst,fig:process-congruence-lfst}).

\section{Translation: Imperative to Functional}
\label{sec:transl-imper-funct}

As a first step, we discuss the translation of the
imperative session type calculus VGR into the linear functional
session type calculus LFST-rec.
The extension with record types is not essential, but it makes
the translation more accessible.  All records could be elided by
replacing them with suitably nested pairs and mapping record labels to
indices.

\subsection{Specification of the Translation}
\label{sec:spec-transl}

The translation from VGR to LFST-rec is type driven, i.e., it is a
translation of typing derivations. The gist of the approach is to translate VGR
expressions into a parameterized linear state transformer monad. It
is parameterized in the sense of Atkey~\cite{DBLP:journals/jfp/Atkey09} because the
type of the state changes with every non-trivial computation step
(i.e., sending and receiving messages).

\begin{figure}[tp]
  \begin{minipage}[t]{0.55\linewidth}
    \begin{align*}
      \intertext{Translation of types}
      \Trans{\Outp T S} &= \Outp{\Trans T} \Trans{ S} \\
      \Trans{\Inp T S} &= \Inp{\Trans T} \Trans{ S} \\
      \Trans\End & = \End \\
      \Trans\TUnit &= \TUnit \\
      \Trans{\Ap S} &= \Ap{\Trans S} \\
      \Trans{\Chan\alpha} &= \TUnit \\
      \Trans{\Sigma_1;T_1 \to T_2; \Sigma_2} &=
                                               \begin{array}[t]{@{}l}
                                               \Trans{T_1} \to %\\
                                               \Record{\Trans{\Sigma_1}}
                                               \to \\ (\Trans{T_2} \times
                                               \Record{\Trans{\Sigma_2}})
                                               \end{array}
      \\
    \end{align*}
  \end{minipage}
  \quad
  \begin{minipage}[t]{0.35\linewidth}
    \begin{align*}
    \intertext{Translation of environments}
    \Trans{\TEnull} &= \Rownull \\
    \Trans{\Sigma, \alpha:T} &= \Trans\Sigma, \alpha:\Trans T \\
    \intertext{Translation of values}
    \Trans{\UnitV} &= \UnitV \\
    \Trans{\gamma^\pm} &= \UnitV \\
    \Trans{x} &= x \\
    \Trans{\Lam x e} &= \Lam x{\Lam \sigma \TransE e} \\
    % \Trans{\Rec x v} &= \Fix\, \Lam x \Trans v \\
    \end{align*}
  \end{minipage}
    \caption{Type translation}
  \label{fig:type-translation-vgr-lfst}
\end{figure}
\begin{figure}[tp]
  \begin{align*}\footnotesize
    \TransE{\ruleVGRApp} &=
                           \begin{array}[t]{@{}l}
                             \Let{(\sigma_1,\sigma_2)}{\Rsplitmany\sigma{\Dom{\Sigma}}} \\
                             \Let{(r, \sigma_1')}{\Trans v\, \Trans{v'}\,\sigma_1} \\
                             (r, \Rconcat{\sigma_1'}{\sigma_2})
                           \end{array}
    \\
    \TransE{
    \ruleVGRReceiveD
    } &=
                        \begin{array}[t]{@{}l}
                        \Let{(c, \sigma')} {\Rsplit \sigma \alpha} \\
                        \Let{(r, c)}{\Receive c} \\
                        (r, \Rconcat{\sigma'}{\Rsingle\alpha c} )
                        \end{array}
    \\
    \TransE{
    \ruleVGRSendD
    } &=
        \begin{array}[t]{@{}l}
          \Let{(c, \sigma')} {\Rsplit\sigma\alpha} \\
          % \Let{(p, \sigma)} {\Rsplit\sigma\beta} \\
          \Let c {\Send {\Trans v}\,c} \\
          (\UnitV, \Rconcat{\sigma'}{\Rsingle\alpha c})
        \end{array}
    \\
    \TransE{
    \ruleVGRReceiveS
    } &=
        \begin{array}[t]{@{}l}
          \Let{(c, \sigma')} {\Rsplit \sigma \alpha} \\
          \Let{(r, c)}{\Receive c} \\
          (\UnitV, \Rconcat{\sigma'}{\Rconcat{\Rsingle d
          r}{\Rsingle\alpha c}})
        \end{array}
    \\
    \TransE{
    \ruleVGRSendS
    } &=
        \begin{array}[t]{@{}l}
          \Let{(c, \sigma')} {\Rsplit\sigma\alpha} \\
          \Let{(p, \sigma'')} {\Rsplit{\sigma'}\beta} \\
          \Let c {\Send p\,c} \\
          (\UnitV, \Rconcat{\sigma''}{\Rsingle\alpha c})
        \end{array}
    \\
    \TransE{
    \ruleVGRClose
    } &=
        \begin{array}[t]{@{}l}
          \Let{(c, \sigma')} {\Rsplit\sigma\alpha} \\
          \Let \_ {\Close c} \\
          (\UnitV, \sigma')
        \end{array}
    \\
    \TransE{
    \ruleVGRLet
    } &=
        \begin{array}[t]{@{}l}
          \Let{(x, \sigma')} {\TransE e}
          \TransE[\sigma'] t
        \end{array}
    \\
    \TransE{
      % \JVGRExpr\Gamma\Sigma{(\Fork  t_1;t_2)}{\Sigma_2}{T_2}{\{\}}
    \ruleVGRForkRevised
    } &=
                                      \begin{array}[t]{@{}l}
                                        \Let{(\sigma_1, \sigma_2)} {\Rsplitmany
                                        \sigma{\Dom{\Sigma_1}}}
                                      \\
                                        \Let {()} {\Fork\TransE[\sigma_1]{t_1}}
                                        \\
                                        \TransE[\sigma_2]{t_2}
                                      \end{array}
    \\
    \TransE{\ruleVGRNew}
    &= {(\New {\Trans S}, \sigma)}
  \end{align*}
  \caption{Translation of expressions and threads (excerpt)}
  \label{fig:expression-translation-vgr-lfst-excerpt}
\end{figure}
\begin{figure}[tp]
  \begin{align*}
    \Trans{\ruleVGRThread} &=
                             \begin{array}[t]{@{}l}
                               \Thread{\Let \sigma
                             {\Rsingle{\vec\gamma}{\vec\gamma}} \TransE t} \\
                               \mathit{where}~ \vec\gamma = \Dom{\Sigma' \setminus \Sigma}
                             \end{array}
    \\
    \Trans{\ruleVGRPar} &= \Trans{C_1} \| \Trans{C_2}
    \\
    \Trans{\ruleVGRNewN} &= (\nu n) \Trans C
    \\
    \Trans{\ruleVGRNewB} &= (\nu \gamma^+\gamma^-) \Trans C
    \\
    \Trans{\ruleVGRNewC} &= (\nu \gamma^+\gamma^-) \Trans C
  \end{align*}
  \caption{Translation of configurations}
  \label{fig:configuration-translation-vgr-lfst-excerpt}
\end{figure}
We map derivations for VGR value typing, VGR
expression typing, and VGR configuration typing to LFST-rec
expressions and configurations.
For brevity, we indicate
the translation with
$\Trans e$ and $\Trans C$  where the arguments are really the
typing derivations for $e$ and $C$, respectively. The translations on
types $\Trans T$, environments $\Trans\Gamma$, $\Trans\Sigma$, and
values $\Trans v$ are homomorphic by induction on the syntax (see
\Cref{fig:type-translation-vgr-lfst}), except for the cases for
channels and functions.
% Channels are mapped to the unit type;
% functions are lifted to the parameterized monad, that is, they
% obtain a state parameter and return an additional state result.

% \begin{align*}
%   \Trans{\Chan\alpha} &= \TUnit &
%                                  \Trans{\gamma^\pm} &= \UnitV \\
%   \Trans{\Sigma_1;T_1 \to T_2; \Sigma_2} &=
%                                            \Trans{T_1} \to %\\
%   \Record{\Trans{\Sigma_1}}
%   \to  (\Trans{T_2} \times
%   \Record{\Trans{\Sigma_2}})
%                       &
%                         \Trans{\Lam x e} &= \Lam x{\Lam \sigma \TransE e} \\
% \end{align*}

The translations are designed to enable proving the following
preservation results.
\begin{prop}[Typing Preserving Translation]\label{thm:translation-preserves-typing}
  \begin{gather*}
    \inferrule[Preserve-Value]
    { \JVGRValue\Gamma v T }
    { \Trans\Gamma \vdash \Trans v : \Trans T }
    \quad
    \inferrule[Preserve-Expression]
    { {\JVGRExpr\Gamma\Sigma e {\Sigma_1}T{\Sigma_2}} }
    {
      \Trans\Gamma, \sigma : \Record{\Trans{\Sigma \setminus \Sigma_1}}
      \vdash \TransE{e} : \Trans T \times
      \Record{\Trans{\Sigma_2}} }
    \quad
    \inferrule[Preserve-Config]{ \JVGRConfig\Gamma\Sigma C {\Sigma_1} }{ \Trans\Gamma,
      {\Trans{\Sigma \setminus \Sigma_1}} \vdash \Trans C }
  \end{gather*}
\end{prop}
\begin{proof}
  See supporting Agda script~\cite{saffrich_hannes_2021_5764782}.
\end{proof}
These statements are proved by mutual induction on the
derivations of the VGR judgments in the premises. The VGR typing judgments for
expressions and configurations pass through unused channels (in $\Sigma_1$) 
in the style of leftover typings~\cite{DBLP:conf/types/Allais17}. 
While this style is convenient for some proofs, it cannot be used for the translation as it fails
when trying to translate the term $\Fork t_1; t_2$. The first premise of its typing rule
\TirName{C-Fork} is $\ruleVGRForkFirstPremise$, which says that executing $t_1$ consumes some of the
incoming channels $\Sigma$ and does not touch the ones in $\Sigma_1$. The second premise
$\ruleVGRForkSecondPremise$ picks up $\Sigma_1$ and demands that $t_2$
consumes all its channels.
However, this pattern does not work for the translation, which is based on
explicit channel passing: if we passed all channels in $\Sigma$ to $t_1$,
which is forked as a new thread, there would be no way to obtain the leftover channels $\Sigma_1$
after thread $t_1$ has finished. Moreover, these channels have to be available for $t_2$ even before $t_1$ has
finished! The same issue arises when translating the parallel
composition of two configurations.
For that reason, in LFST-rec the translated expressions and configurations are
supplied with exactly the channels needed.

\Cref{fig:type-translation-vgr-lfst} contains the details of the type translation, the
translation of environments, and the translation of values. The only interesting case of the type
translation is the one for function types, which maps a function to a Kleisli arrow in a linear,
parameterized state monad.
The incoming and outgoing channel environments are mapped to the incoming and outgoing state record
types.  The other observation is that any channel type is mapped to the unit type.

The translation of values has two interesting cases. A channel value
is mapped to the unit value $\UnitV$ because
channels are handled on the type level and channel references are resolved by
accessing the corresponding field of the state record. Functions obtain an extra argument $\sigma$ for
the incoming state record that contains the currently open
channels. The body of a lambda is translated by the expression
translation which is indexed by
the incoming state record $\sigma$ and returns a pair of the result
and the outgoing state record.

\Cref{fig:expression-translation-vgr-lfst-excerpt}
shows select cases from the translation of expressions  that demonstrate the role of the record
operations. The conclusion of
\TirName{Preserve-Expression} shows that an expression is correctly translated
to a linear state transformer as in the translation of the
function type.

\Cref{fig:configuration-translation-vgr-lfst-excerpt} contains
the translation of the configuration typing rules. Of those, the most
interesting case is the \TirName{C-Thread} configuration rule. Threads
execute in a context that contains a list of access points with their
types. The thread body may refer to channels in $\Sigma$. The
translation rule
reifies the channels that are used in the thread by collecting them in a record $\sigma$ and
injecting that record as the initial state of the state monad. This record is transformed by the
expression translation that returns a pair of the return value of type $\Trans T$ and the final
record of type $\Record{\Trans{\TEnull}}$. It is easy to
see that this pair is unrestricted because the translation of a (non-session) type $T$ is generally
unrestricted and the empty record is also unrestricted.

The remaining rules are simple. In \TirName{C-Par}, we do not have to
manipulate the channel environments as we do in the \TirName{T-Fork}
rule because channels are only reified at the thread level in rule
\TirName{C-Thread}. \TirName{C-NewN} creates a new access point,
\TirName{C-NewB} compensates for the different handling of channel
restriction in VGR and LFST-rec. \TirName{C-NewC} handles depleted
channels.

\subsection{Simulation}
\label{sec:results}

We would like the translation to induce a simulation in that
each step of a typed VGR configuration $C$ gives rise to one or more
steps in its translation $\Trans C$ in LFST-rec. Unfortunately, the
situation is not that simple because administrative reductions
involving the state get in the way.
\begin{prop}[Simulation]\label{proposition:simulation}
  If $\JVGRConfig\Gamma\Sigma C{\Sigma'}$
  and $C \VGRPReduceTo[\RLabel] C'$ in VGR,
  then there is a configuration $\Adorn C$ in LFST-rec such that
  $\Trans C \GVPReducePlus[\RLabel] \Adorn C$ and
  $\Trans{C'} \GVPReducePlus[\ELabel] \Adorn C$.
\end{prop}
\begin{proof}
  See appendix \Cref{sec:proof-prop-simulation}.
\end{proof}

\section{Translation: Functional to Imperative}
\label{sec:backw-transl}

For the backwards translation we consider LFST programs without
records and we informally extend the
expression language of VGR with pairs---analogous to LFST, but unrestricted.

We first define an untyped translation that demonstrates that the
calculi are equally expressive. Then we define a restricted version of
LFST's type system to characterize the subset of LFST on which the translation preserves typing.

\subsection{Untyped Translation}
\label{sec:untyped-translation}

In a first approximation,
the backwards translation, indicated by $\Back{e}$ for an LFST
expression $e$,  might map the send and receive operations
naively as follows.
\begin{align}
  \Back{\Send {e_1}{e_2}}
  &=
                            \Let x {\Back{e_1}} \Let y {\Back{e_2}}
                            \Let z {\Send x y} y
  \\\label{eq:22}
  \Back{\Receive e} &=\Let y {\Back e} \Let x {\Receive y} (x,y)
\end{align}
This mapping, extended analogously to the rest of LFST, yields a program
in A-normal form that fits with VGR's syntactic
restrictions. The functional send operation returns the updated
channel, so we have to duplicate the channel reference $y$ in its image in
VGR\@. Similarly, the functional receive operation returns a pair of the
received value and the updated channel, so the translation needs to
construct a pair from the  received value and the updated channel
$y$.

However, to prove a tight relation between reduction in LFST and VGR,
we need to be more careful to avoid administrative reductions.
For example, if $e$ in~\eqref{eq:22} is already a value, then the
inserted $\Let y {\Back e} \dots$ is gratuitous and results in an
extra administrative reduction in VGR\@.

This phenomenon is known since Plotkin's treatise of the
CPS translation~\cite{DBLP:journals/tcs/Plotkin75}.
Hence, we factor the backwards translation in two steps. The first step
transforms the LFST program to A-normal form using an approach due to
Sabry and Felleisen~\cite{DBLP:journals/lisp/SabryF93}. This
transformation is known to give rise to a strong operational
correspondence (a reduction correspondence~\cite{DBLP:journals/toplas/SabryW97}), it is
typing preserving, and it is applicable to LFST because it preserves
linearity. The definition of this translation $\TOANF e$ is given in \Cref{sec:translation-anf}.

This refined ANF translation is compatible with evaluation because it is compatible
with values, evaluation contexts, and substitution.
\begin{lem}[Value preservation]\label{lemma:anf-value-preservation}
  $\TOANF v$  is a value in LFST\@.
\end{lem}
\begin{proof}
  Case analysis on values $v$.
\end{proof}
\begin{lem}[Evaluation preservation]\label{lemma:anf-evaluation-preservation}
  If $E$ is an LFST evaluation context, then so is $\TOANF E$.
\end{lem}
\begin{proof}
  Induction on the definition of evaluation contexts
  (\Cref{fig:semantics-lfst}) using the definition of the ANF
  translation for   evaluation contexts
  (\Cref{fig:lfst-translation-anf-evaluation-context}).
\end{proof}
\begin{lem}[Substitution]\label{lemma:anf-substitution-preservation}
  $\TOANF{e}[ \TOANF{v} / x]  = \TOANF{e[v/x]}$.
\end{lem}
\begin{proof}
  Induction on $e$. The only interesting case arises for $e=x$:
  \begin{gather*}
    \TOANF x[\TOANF v /x] = x[\TOANF v /x] = \TOANF v = \TOANF{x[v/x]}
  \end{gather*}
  All other cases are immediate by the induction hypothesis.
\end{proof}

\begin{prop}[ANF Simulation]\label{lemma:anf-simulation}
  \begin{enumerate}
  \item If $e \GVEReduceTo e'$, then $\TOANF e \GVEReducePlus
    \TOANF{e'}$.
  \item If $C \GVPReduceTo[\RLabel] C'$, then $\TOANF C \GVPReducePlus[\RLabel] \TOANF{C'}$.
  \end{enumerate}
\end{prop}
\begin{proof}
  See \Cref{sec:proof-prop-anf-simulation}.
\end{proof}

The second step is the expression
translation $\Back e$ from LFST-ANF to VGR\@. This translation is very
simple because the source calculus is already in A-normal form. The
idea of the translation as stated at the beginning of this section is
clearly reflected in the first two lines of the expression translation
$\Back e$. The remaining cases of the translation proceed homomorphically (see
\Cref{fig:translation-lfst-anf-vgr} in \Cref{sec:expr-transl}).
\begin{align*}
      \Back{\Send {v}{w}}
      &=
        \Let z {\Send {\Back{v}} {\Back{w}}} {\Back{w}}
      \\
      \Back{\Receive v} &=\Let x {\Receive {\Back v}} (x,\Back v)
      \\
      \Back{\Fork e} &= {\Fork{\Back e}}; \UnitV
\end{align*}

This setup establishes a tight connection between LFST-ANF and VGR,
because the translation preserves values, evaluation contexts,
and substitution.
\begin{lem}[Value preservation]\label{lemma:value-preservation}
  For each value $v$ of LFST-ANF, $\Back v$  is a VGR value.
\end{lem}
\begin{proof}
  Simple case analysis.
\end{proof}
\begin{lem}[Evaluation preservation]\label{lemma:evaluation-preservation}
  \begin{enumerate}\item
For each evaluation context $E$ of LFST-ANF, $\Back E$ is a VGR
    evaluation context.
  \item
    For each expression $e$ of LFST-ANF,
    $\Back{E[e]} = \Back E[\Back e]$.
  \end{enumerate}
\end{lem}
\begin{proof}
  In LFST-ANF, the grammar of evaluation contexts is reduced to
  \begin{align*}
    E & ::= \Hole \mid \Let x E e
  \end{align*}
  which clearly matches VGR evaluation contexts (cf.\
  \Cref{fig:semantics-vgr}). So, item~1 is immediate and item~2 holds
  by induction on $E$.
\end{proof}
\begin{lem}[Backwards substitution]\label{lemma:substitution-preservation}
  For each LFST-ANF expression $e$ and value $v$,
  $\Back{e}[ \Back{v} / x]  = \Back{e[v/x]}$.
\end{lem}
\begin{proof}
  Induction on $e$ using \Cref{lemma:value-preservation} for
  the case $e=x$.
\end{proof}

\begin{prop}[Backwards simulation]\label{lemma:back-simulation}
  Let $e, e'$ and $C,C'$ be expressions and configurations in LFST-ANF\@.
  \begin{enumerate}
  \item If $e \GVEReduceTo e'$, then $\Back e \VGREReduceTo
    \Back{e'}$.
    \qquad
  \item If $C \GVPReduceTo[\RLabel] C'$, then $\Back C \VGRPReducePlus[\RLabel] \Back{C'}$.
  \end{enumerate}
\end{prop}
\begin{proof}
  See \Cref{sec:proof-prop-backwards-simulation}.
\end{proof}

Putting the results for the two steps together, we obtain the desired
tight simulation result by composing \Cref{lemma:anf-simulation,lemma:back-simulation}.
\begin{prop}[Full Backwards Simulation]
  Suppose that $e, e'$ and $C,C'$ are expressions and configurations
  in LFST\@.
  \begin{enumerate}
  \item If $e \GVEReduceTo e'$, then $\Back{\TOANF e} \VGREReducePlus
    \Back{\TOANF{e'}} $.
    \qquad
  \item If $C \GVPReduceTo[\RLabel] C'$, then $\Back{\TOANF C}
    \VGRPReducePlus[\RLabel] \Back{\TOANF{C'}}$.
  \end{enumerate}
\end{prop}

\subsection{Typed Backwards Translation}
\label{sec:towards-typed-backw}

To obtain a type preserving backwards translation from LFST to VGR, we
have to add extra information to the type system of
LFST\@. Unfortunately, this extra information makes the typing more
restrictive. We start with an informal review of the requirements.

First, as VGR tracks channel identities, they have to be
represented in the revised type system for LFST\@. 
Following Padovani~\cite{DBLP:conf/esop/Padovani17}, 
we tag session types as in $\Tagged\alpha s$  consisting of  a session
type $s$ tagged with an identity $\alpha$.
This change affects the following five preliminary typing rules: accept and request
create new channel identities, sending and receiving continues on
the same channel.
\begin{mathpar}
  \ruleGVsendID

  \ruleGVrecvID

  \ruleGVacceptID

  \ruleGVrequestID

  \ruleGVcloseID
\end{mathpar}

Second, the function type in VGR specifies a transformation on the
channels that are implicitly or explicitly affected by the
function. Hence, we must augment the LFST
type system with tracking the identities of channels, on which the
program performs an effect.  To this end, we equip LFST with a
suitable sequential effect system~\cite{DBLP:conf/ecoop/Gordon17}.
It
distinguishes between incoming and outgoing channels, $\Sigma_i$ and
$\Sigma_o$, which are also reflected in the
latent effect on the function arrow.

Hence, the resulting typing judgment
\begin{gather*}
  \JGVExprID\Gamma e {t/\Sigma_i\mapsto\Sigma_o}
\end{gather*}
reads like this: in typing environment $\Gamma$, expression $e$ has
type $t$ and its evaluation processes channels according to $\Sigma_i$
and returns channels according to $\Sigma_o$.

We define tagged session types by adding an identity tag $\alpha$ to all
session types and augmenting function types with a set of uniquely
tagged sessions. We carve out a set of data types $d$, which can be transmitted in VGR programs. Hence,
session types proper (denoted by $s$) are a subset of LFST's session types.
\begin{align*}
  & \mathrm{Types}&
                    t &::=  \Tagged \alpha s \mid \Ap s \mid \TUnit
                        \mid t \to^{\Sigma\mapsto\Sigma} t \mid t
                        \linto^{\Sigma\mapsto\Sigma} t \mid t\lpair t
  \\
  & \mathrm{Data} &
                    d &::= \Ap s \mid \TUnit \mid  t \to^{\Sigma\mapsto\Sigma} t \mid t
                        \linto^{\Sigma\mapsto\Sigma} t
  \\
  & \mathrm{Sessions} &
                        s & ::= \Inp d s \mid \Outp d s \mid \Inp s s
                            \mid \Outp s s \mid \End
\end{align*}
\begin{figure}[tp]
  \input{mathpar-gv-eff-typing-excerpt}
  \caption{Typing rules for LFST-EFF (excerpt)}
  \label{fig:lfst-eff-typing-excerpt}
\end{figure}
\begin{figure}[tp]
  \begin{minipage}[t]{0.45\linewidth}
\begin{align*}
    \Erase{\Tagged\alpha s} & = \Erase s \\
    \Erase{\Ap s} &= \Ap{\Erase s} \\
    \Erase{\TUnit} &= \TUnit \\
    \Erase{t \to^{\Sigma\mapsto\Sigma'} t'} &= \Erase t \to \Erase{t'}
    \\
    \Erase{t \linto^{\Sigma\mapsto\Sigma'} t'} &= \Erase t \linto
                                                 \Erase{t'} \\
    \Erase{ t\lpair t'} &= \Erase t \lpair \Erase{t'}
  \end{align*}
  \end{minipage}
  \quad
  \begin{minipage}[t]{0.45\linewidth}
    \begin{align*}
    \Erase{ \Inp d s} & = \Inp{\Erase d}{\Erase s} \\
    \Erase{ \Outp d s} &= \Outp{\Erase d}{\Erase s} \\
    \Erase{ \Inp s s'} &= \Inp{\Erase s}{\Erase{s'}} \\
    \Erase{ \Outp s s'} &= \Outp{\Erase s}{\Erase{s'}}\\
    \Erase\End &= \End
    \end{align*}
  \end{minipage}
  \caption{Effect type erasure}
  \label{fig:effect-type-erasure}
\end{figure}
Using mostly standard effect typing rules (see
\Cref{fig:lfst-eff-typing-excerpt} for select rules and
\Cref{sec:effect-typing} for the full set of rules),
we show that effect typing is a proper restriction of LFST typing.

We write $\Erase t$ for the erasure of  an LFST-EFF type
$t$, which is defined in \Cref{fig:effect-type-erasure}. Erasure
extends pointwise to environments $\Gamma$.
\begin{lem}[Conservative Extension]
  $\JGVExpr[']\Gamma e {t /\Sigma\mapsto\Sigma'}$ implies $\JGVExpr{\Erase\Gamma}  e {\Erase t}$.
\end{lem}
\begin{proof}
  Straightforward induction. The standard typing rules correspond to
  the erasure of the effect typing rules.
\end{proof}
The translation to ANF does not affect LFST typing with effects.
\begin{lem}[ANF Compatible]\label{lemma:anf-compatible}
  Suppose that $\JGVExpr[']\Gamma e {t / \Sigma \mapsto \Sigma'}$.

  Then  $\JGVExpr[']\Gamma {\TOANF e} {t / \Sigma \mapsto \Sigma'}$.
\end{lem}
\begin{proof}
  See \Cref{sec:proof-lemma-anf-compatible}.
\end{proof}

\Cref{fig:lfst-eff-to-vgr-types} contains the backwards
translation for types. An $\alpha$-tagged
session type turns into the channel type $\Chan\alpha$ and the effect
annotation on function types gets mapped to the before and after
environments in VGR function types.
\begin{figure}[tp]
  \begin{minipage}[t]{0.22\linewidth}
    \begin{align*}
      \Back{\Outp t s} &= \Outp{\Back t}{\Back s}  \\
      \Back{\Inp t s} &= \Inp{\Back t}{\Back s}  \\
      \Back{\End} &= \End \\
    \end{align*}
  \end{minipage}
  \begin{minipage}[t]{0.22\linewidth}
    \begin{align*}
      \Back{\Ap s} &= \Ap{\Back s} \\
      \Back{\Tagged \alpha s} &= \Chan \alpha \\
      \Back{\TUnit} &= \TUnit \\
    \end{align*}
  \end{minipage}
 \begin{minipage}[t]{0.5\linewidth}
   \begin{align*}
     \Back{t_1 \to^{\Sigma_1\mapsto\Sigma_2} t_2} &= \Sigma_1; {\Back{t_1}} \to \Back{t_2}; \Sigma_2\\
     \Back{t_1 \linto^{\Sigma_1\mapsto\Sigma_2} t_2} &= \Sigma_1; \Back{t_1} \to\Back{t_2}; \Sigma_2\\
     \Back{t_1 \otimes t_2} &= \Back{t_1} \times \Back{t_2}
   \end{align*}
 \end{minipage}
%    \vspace{-\baselineskip}
  \caption{Type translation from LFST-EFF to VGR}
  \label{fig:lfst-eff-to-vgr-types}
\end{figure}

This preparation enables us to prove the typing preservation of the
backwards translation.
\begin{prop}[Typing Preservation (Backwards)]\label{proposition:typing-preservation-backwards}~\\
  Suppose that $\JGVExpr[']\Gamma e {t/\Sigma_1\mapsto\Sigma_2}$ is an
  LFST-EFF typing for   some expression $e$ in LFST-ANF\@.

  For all $\Sigma$ such that $\Sigma\#\Sigma_1$ and $\Sigma\#\Sigma_2$,
  $\JVGRExpr {\Back\Gamma} {\Sigma, \Sigma_1} {\Back e} {\Back t} {\Sigma} {\Sigma_2}$.
\end{prop}
\begin{proof}
  See \Cref{sec:proof-prop-typing-preservation-backwards}.
\end{proof}

\section{Related Work}
\label{sec:related-work}

Pucella and Tov~\cite{DBLP:conf/haskell/PucellaT08} give an embedding of a session
type calculus in Haskell. Like our translation, their embedding relies
on a parameterized monad, which is layered on top of the IO monad
using phantom types. Linearity is enforced by the monad abstraction.
Multiple channels are implemented by stacking so that channel names
are de Bruijn indices. Stacking only happens at the (phantom) type level,
so that stack rearrangement has no operational consequences. The paper
comes with a formalization and a soundness proof of the implementation.
Sackman and Eisenbach~\cite{SackmanE08} also encode session types for a single channel in
Haskell using an indexed (parameterized) monad.

Imai and coworkers~\cite{DBLP:journals/scp/ImaiYY19} propose an
encoding of binary session-based communication as a library in
OCaml. This library is based on an indexed state monad that maintains the
current state of a set of channels in a tuple. Channel names are
encoded by lenses operating on this state and operations an a channel change the
index type at the position indicated by the lens. The programming
style resembles VGR, but it is explicitly monadic. The monad and its
type indexing are closely related to our encoding, which is linear by
typing.

Another line of work on session types is based on process calculi
obtained through the Curry-Howard correspondence applied to fragments
of linear logic~\cite{DBLP:conf/concur/CairesP10,DBLP:conf/esop/BalzerTP19,DBLP:conf/concur/DasP20}. 
The resulting programs have an imperative flavor as they are based on
process calculus. The correspondence structures communication as a
string of interactions on a channel name. This channel name ``changes
type'' by rebinding at each communication operation. There is a
monadic embedding of this approach into a pure functional language~\cite{DBLP:conf/esop/ToninhoCP13}. 
In this stratified language, processes are snippets of
imperative code encapsulated as first-class monadic values into the functional
language. These values can be plugged into a process term by a
suitable adaptation of the monadic bind operation. Processes may transmit
channel names or values from the functional stratum. Processes have
the imperative flavor as already mentioned. It would be interesting
future work to relate this line of work with the correspondence
developed in the present paper.

Alias types~\cite{DBLP:conf/esop/SmithWM00} presents
a type system for a low-level language where the type of a function expresses the shape of the store
on which the function operates. Function types can abstract over store locations
$\alpha$ and the shape of the store is described by \emph{aliasing
  constraints} of the form $\{\alpha\mapsto T\}$. Constraint
composition resembles separating conjunction~\cite{DBLP:conf/lics/Reynolds02} 
and ensures that locations are
unique. Analogous to our channel types, pointers in the
alias types system can be duplicated and have a singleton type
indicating their store location. Alias types also
include non-linear constraints, which are not required in our system.

\section{Conclusion}
\label{sec:conclusion}

Disregarding types, the imperative and functional session
calculi are equally powerful. But typing is the essence of a session
calculus so that the imperative calculus is strictly less
expressive. Two issues are responsible for the limitations.
\begin{enumerate}
\item Identity tracking for channels restricts the usability of
  functional abstraction. As soon as types represent channel
  identities, functions are fixed to specific channels in a simply
  typed system.
  % and the creation of channels is limited
\item Having different typing rules for sending channels and sending
  (other) data impedes abstraction and
  modularity. Higher-order channel passing has subtle
  problems that limit the usefulness of a transmitted channel.
\end{enumerate}

Our results suggest that the simple nature of VGR's type system is the
culprit for the severe restrictions on expressiveness. On the other
hand, the conciseness of VGR programs is appealing to many (imperative) programmers.
Hence, it is an interesting future work to  extend VGR's type system such that there are
type and semantics preserving translations in both directions.
As demonstrated by the work on Alias Types~\cite{DBLP:conf/esop/SmithWM00}, 
polymorphism over identities is one
required ingredient, but more work is needed to clarify all issues
involved in such a system.

\clearpage
%
% ---- Bibliography ----
%
\bibliographystyle{alphaurl}
\bibliography{all}
% \end{document}

%% Appendix
\clearpage
\appendix
\section{Appendix}

% \subsection{VGR}
% \label{sec:typing-vgr}
% \input{fig-vgr-expression-typing}
% \input{fig-vgr-configuration-typing}

\subsection{LFST}
\label{sec:typing-lfst}

% \begin{figure}[tp]
%   \input{mathpar-gv-unrestricted-types}
%   LFST expression typing \hfill \fbox{$\JGVExpr{\Gamma} e t$}
%   \begin{mathpar}
%     \ruleGVunit
%
%     \ruleGVvar
%
%     \ruleGVlamunr
%
%     \ruleGVlamlin
%
%     \ruleGVapp
%
%     \ruleGVpair
%
%     \ruleGVpairE
%   \end{mathpar}
%   \begin{mathpar}
%     \ruleGVfork
%
%     \ruleGVsend
%
%     \ruleGVrecv
%
%     \ruleGVnew
%
%     \ruleGVaccept
%
%     \ruleGVrequest
%   \end{mathpar}
%
%   \input{mathpar-gv-record-typing-rules}
%   \caption{Typing rules of LFST}
%   \label{fig:typing-lfst}
% \end{figure}
\begin{figure}[t!]
  \begin{align*}
    & \text{Values} &
    v,w &::= x \mid \UnitV \mid \Lam x e \mid (v, w) \mid
          \Record{\vec\alpha=\vec v} \\
    & \text{Evaluation contexts} &
                                   E &::= \Hole \mid E\,e \mid v\,E
                                       \mid (E, e) \mid (v, E) \mid
                                       \Let{(x,y)} E e
    \\&&&
                                       \mid \Send E e \mid \Send v E
                                       \mid \Receive E
                                       \mid \Accept E \mid \Request E
                                       \mid \Close E
    \\&&&
                                       \mid
                                       \Record{\vec\alpha=\vec
                                       v, \alpha= E,
          \vec\beta=\vec e}
          \mid \Rsplitmany E {\vec\alpha}
          \mid \Rconcat E e \mid \Rconcat v E
          \mid \Rsplit E \alpha
    \\& \text{Processes} &
                           C, D & ::= \Thread e \mid C \| D \mid
                                  (\nu\gamma\delta)~C \mid (\nu p)~C
  \end{align*}
  LFST expression reduction
  \begin{mathpar}
    \inferrule{}{(\Lam x e)\,v \GVEReduceTo e[v/x]}

    \inferrule{}{\Let{(x,y)}{(v,w)} e \GVEReduceTo e[v,w/x,y]}

    \inferrule{}{
      \Rsplit {
        \Record{\alpha=v, \vec\beta=\vec w}
      } \alpha \GVEReduceTo
      (v, \Record{\vec\beta=\vec w})
    }

    \inferrule{}{
      \Rsplitmany {
        \Record{\vec\alpha=\vec v, \vec\beta=\vec w}
      } {\vec\alpha} \GVEReduceTo
      (\Record{\vec\alpha=\vec v},
      \Record{\vec\beta=\vec w} )
    }

    \inferrule{
      \vec\alpha \# \vec\beta
    }{
      \Rconcat{\Record{\vec\alpha=\vec v}}{
        \Record{\vec\beta=\vec w}}
      \GVEReduceTo
      \Record{\vec\alpha=\vec v, \vec\beta=\vec w}
    }
  \end{mathpar}

  LFST process reduction (in any process context and modulo process congruence)
  \begin{mathpar}
    \inferrule{ e \GVEReduceTo e' }{
      \Thread{E[e]} \GVPReduceTo
      \Thread{E[e']}}

    \inferrule{}{
      \Thread{E[\Fork e]} \GVPReduceTo[\KWFork] \Thread{E[\UnitV]}
      \| \Thread e
    }

    \inferrule{}{
      \Thread{E[\New s]} \GVPReduceTo[\KWNew] (\nu p) \Thread{E[p]}
    }

    \inferrule{}{
      \Thread{E[\Accept p]} \| \Thread{F[\Request p]}
      \GVPReduceTo[\KWAccept]
      (\nu\gamma\delta) (\Thread{E[\gamma]} \| \Thread{F[\delta]})
    }

    \inferrule{}{
      (\nu\gamma\delta) (\Thread{E[\Send v \gamma]} \| \Thread{F[\Receive \delta]})
      \GVPReduceTo[\KWSend]
      (\nu\gamma\delta) (\Thread{E[\gamma]} \| \Thread{F[(v, \delta)]})
    }

    \inferrule{}{
      (\nu\gamma\delta) (\Thread{E[\Close \gamma]} \| \Thread{F[\Close \delta]})
      \GVPReduceTo[\KWClose]
      \Thread{E[\UnitV]} \| \Thread{F[\UnitV]})
    }
  \end{mathpar}
  \caption{Semantics of LFST}
  \label{fig:semantics-lfst}

  \begin{mathpar}
    \inferrule{}{ C \| C' \Pcong C' \| C}

    \inferrule{}{
      (C \| C') \| C'' \Pcong C \|(C' \| C'')
    }

    \inferrule{}{
      (\nu\gamma\delta)~C \Pcong (\nu\delta\gamma)~C
    }

    \inferrule{\delta,\gamma\notin C}{
      C \| (\nu\gamma\delta)~ C' \Pcong  (\nu\gamma\delta)( C \| C')
    }

    \inferrule{p\notin C}{
      C \| (\nu p)~ C' \Pcong  (\nu p)( C \| C')
    }
  \end{mathpar}
  \caption{Process congruence in LFST}
  \label{fig:process-congruence-lfst}
\end{figure}

\Cref{fig:semantics-lfst} defines a synchronous version of the dynamic semantics of
LFST~\cite{DBLP:journals/jfp/GayV10}, which is similar to the
semantics of context-free session types~\cite{DBLP:conf/icfp/ThiemannV16}.
\Cref{fig:process-congruence-lfst} defines the standard process
congruence rules that make processes into a commutative semigroup.

\subsection{Translation to ANF}
\label{sec:translation-anf}

\Cref{fig:lfst-translation-anf}
contains the definition of
the translation of LFST to LFST in A-normal form. It is defined by induction on LFST expressions with
the additional twist that it distinguishes between non-value terms $n$
and value terms $v$ (see \Cref{fig:semantics-lfst}). The idea is that intermediate let-expressions are
only introduced if the current term is a non-value. Variables that
only appear on the right hand side in the translation are assumed to
be fresh.

Formally, we define non-values in LFST as follows:
\begin{align*}
  n &::= n\,e \mid e\,n \mid (n, e) \mid (e, n) \mid \Let{(x,y)}e e \\
  &\mid \Send {e}{e} \mid \Receive e \mid \Accept e \mid \Request e \mid \Close e
\end{align*}
\begin{figure}[tp]
    \begin{align*}
      \intertext{Value translation $\TOANF{v}$}
      \TOANF{x} &= x
      \\
      \TOANF{\UnitV} &= \UnitV
      \\
      \TOANF{\Lam x e}&=\Lam x \TOANF{e}
      \\
      \TOANF{n\,e}&= \Let x{\TOANF{n}} \TOANF{x\,e}
      \\
      \TOANF{v\,n}&= \Let y{\TOANF{n}} {\TOANF{v}\,y}
      \\
      \TOANF{v\,w}&= {\TOANF{v}\,\TOANF{w}}
      \\
      \TOANF{(n, e)}&= \Let x{\TOANF{n}} {\TOANF{(x, e)}}
      \\
      \TOANF{(v, n)}&= \Let y{\TOANF{n}} {(\TOANF{v}, y)}
      \\
      \TOANF{(v, w)}&= {(\TOANF{v},\TOANF{ w})}
      \\
      \TOANF{\Let{(x,y)}{n}{e}} &= \Let z {\TOANF n}\Let{(x,y)}{z}{\TOANF{e_2}}
      \\
      \TOANF{\Let{(x,y)}{v}{e}} &= \Let{(x,y)}{\TOANF{v}}{\TOANF{e}}
      \intertext{Expression translation $\TOANF{e}$}
      \TOANF{\Send {n}{e}}
      &=
        \Let x {\TOANF{n}} {\TOANF{\Send x {e}}}
      \\
      \TOANF{\Send {v}{n}}
      &=
        \Let y {\TOANF{n}} {\Send {\TOANF{v}} y}
      \\
      \TOANF{\Send {v}{w}}
      &=
        {\Send {\TOANF{v}} {\TOANF{w}}}
      \\
      \TOANF{\Receive n} &=\Let y {\TOANF n}  {\Receive y}
      \\
      \TOANF{\Receive v} &= {\Receive {\TOANF v}}
      \\
      \TOANF{\Close n} &=\Let y {\TOANF n}  {\Close y}
      \\
      \TOANF{\Close v} &= {\Close {\TOANF v}}
      \\
      \TOANF{\Accept n} &= \Let x{\TOANF n} {\Accept x}
      \\
      \TOANF{\Accept v} &= {\Accept {\TOANF v}}
      \\
      \TOANF{\Request n} &= \Let x{\TOANF n}{\Request x}
      \\
      \TOANF{\Request v} &= {\Request{\TOANF v}}
      \\
      \TOANF{\New s} &= \New{\TOANF s}
      \\
      \TOANF{\Fork e} &= {\Fork{\TOANF e}}; \UnitV
                                  \intertext{Process translation $\TOANF
                                  C$}
                                  \TOANF{\Thread e} &= \Thread{\TOANF e}
      \\
      \TOANF{C \| D} &= \TOANF C \| \TOANF D
      \\
      \TOANF{(\nu p)~C} &= (\nu p)~\TOANF C
      \\
      \TOANF{(\nu\gamma^+\gamma^-)~ C} &= (\nu\gamma^+\gamma^-)~\TOANF C
    \end{align*}
  \caption{Translation to A-normal form (LFST)}
  \label{fig:lfst-translation-anf}
\end{figure}
We extend the translation to evaluation contexts as shown in
\Cref{fig:lfst-translation-anf-evaluation-context}.
\begin{figure}[tp]
  \begin{align*}
    \TOANF\Hole &= \Hole \\
    \TOANF{E\,e} &= \Let x{\TOANF E} \TOANF{x\,e} \\
    \TOANF{v\,E} &= \Let y{\TOANF E} \TOANF{v}\,y \\
    \TOANF{(E,e)} &= \Let x{\TOANF E} \TOANF{(x,e)} \\
    \TOANF{(v,E)} &= \Let y{\TOANF E} (\TOANF{v},y) \\
    \TOANF{\Let{(x,y)} E e} &= \Let z{\TOANF E} \Let{(x,y)} z \TOANF e \\
    \TOANF{\Send Ee} &= \Let x{\TOANF E} \TOANF{\Send xe} \\
    \TOANF{\Send vE} &= \Let y{\TOANF E} \Send{\TOANF{v}}y \\
    \TOANF{\Receive E} &= \Let y {\TOANF E} \Receive y \\
    \TOANF{\Close E} &= \Let y {\TOANF E} \Close y \\
    \TOANF{\Accept E} &= \Let y {\TOANF E} \Accept y \\
    \TOANF{\Request E} &= \Let y {\TOANF E } \Request y
  \end{align*}
  \caption{Translation to A-normal form (LFST evaluation contexts)}
  \label{fig:lfst-translation-anf-evaluation-context}
\end{figure}

\subsection{Expression translation}
\label{sec:expr-transl}

\Cref{fig:translation-lfst-anf-vgr} contains  the expression
translation $\Back e$ from LFST-ANF to VGR\@.
\begin{figure}[tp]
    \begin{align*}
      \intertext{Value translation $\Back v$}
      \Back{x} &= x
      \\
      \Back{\UnitV} &= \UnitV
      \\
      \Back{\Lam x e}&=\Lam{(\_; x:\_)}\Back{e}
      \\
      \Back{v\,w}&= {\Back{v}\,\Back{w}}
      \\
      \Back{(v, w)}&= {(\Back{v},\Back{ w})}
      \\
      \Back{\Let{(x,y)}{v}{e}} &= \Let{(x,y)}{\Back{v}}{\Back{e}}
      \\
      \Back{\Let x {e_1} {e_2}} & = \Let x{\Back{e_1}} \Back{e_2}
      \intertext{Expression translation $\Back{e}$}
      \Back{\Send {v}{w}}
      &=
        \Let z {\Send {\Back{v}} {\Back{w}}} {\Back{w}}
      \\
      \Back{\Receive v} &=\Let x {\Receive {\Back v}} (x,\Back v)
      \\
      \Back{\Accept v} &= {\Accept {\Back v}}
      \\
      \Back{\Request v} &= {\Request{\Back v}}
      \\
      \Back{\Close v} &= {\Close{\Back v}}
      \\
      \Back{\New s} &= \New{\Back s}
      \\
      \Back{\Fork e} &= {\Fork{\Back e}}; \UnitV
                                  \intertext{Process translation $\Back C$}
                                  \Back{\Thread e} &= \Thread{\Back e}
      \\
      \Back{C \| D} &= \Back C \| \Back D
      \\
      \Back{(\nu p)~C} &= (\nu p)~\Back C
      \\
      \Back{(\nu\gamma^+\gamma^-)~ C} &= (\nu\gamma)~\Back C
    \end{align*}
  \caption{Translation from LFST-ANF to VGR}
  \label{fig:translation-lfst-anf-vgr}
\end{figure}

\subsection{Effect typing}
\label{sec:effect-typing}

\Cref{fig:lfst-eff-typing} shows a selection of the effect typing rules as the addition of effects is
mostly standard.
\begin{figure}[tp]
  \input{mathpar-gv-eff-typing}
  \caption{Typing rules for LFST-EFF}
  \label{fig:lfst-eff-typing}
\end{figure}

\subsection{Proofs}
\label{sec:proofs}

\subsubsection{Proof of \Cref{proposition:simulation}}
\label{sec:proof-prop-simulation}

\begin{proof} (Sketch)
  Consider a thread reducing $\Thread t \VGRPReduceTo \Thread{t'}$.

  \begin{align*}
    \Trans{\Thread t} & = \Thread{\Let\sigma {\Rsingle{\vec\gamma}{\vec\gamma}}{ \TransE t}} &
    &\GVPReduceTo \Thread{\TransE[{\Rsingle{\vec\gamma}{\vec\gamma}}] t}
    % \\
    % \Trans{\Thread {t'}} & = \Thread{\Let\sigma {\Rsingle{\vec\gamma}{\vec\gamma}}{ \TransE {t'}}} &
    % &\ReduceTo \Thread{\TransE [{\Rsingle{\vec\gamma}{\vec\gamma}}]{{t'}}}
  \end{align*}
  If $t = v\,v'$, then $v = \Lam x e$ and $t' = e[v'/x]$. All
  reductions happen at the expression level.
  \begin{align*}
    \TransE [{\Rsingle{\vec\gamma}{\vec\gamma}}]{(\Lam x e)\, v'}
    &= \Trans{\Lam x e}\,\Trans{v'}\, {\Rsingle{\vec\gamma}{\vec\gamma}} \\
    &= (\Lam x \Lam \sigma \TransE e)\, \Trans{v'}\, {\Rsingle{\vec\gamma}{\vec\gamma}} \\
    &\GVEReduceTo (\Lam\sigma \TransE e [\Trans{v'}/x]) \, {\Rsingle{\vec\gamma}{\vec\gamma}} \\
    &= (\Lam\sigma \TransE {e[v'/x]}) \, {\Rsingle{\vec\gamma}{\vec\gamma}} \\
    &\GVEReduceTo \TransE [ {\Rsingle{\vec\gamma}{\vec\gamma}}] {e[v'/x]} \\
    &= \TransE [{\Rsingle{\vec\gamma}{\vec\gamma}}]{{t'}}
  \end{align*}
  Reduction of $\Let x v t$ is similar.

  If $t= \Receive v$, then $t' = v'$ if $t \VGREReduceTo[{\gamma_0?v'}] t'$.
  \begin{align*}
    \TransE [{\Rsingle{\alpha = \gamma_0, \vec\gamma}{\vec\gamma}}]{\Receive v}
    &= \begin{array}[t]{@{}l}
                         \Let{(c, \sigma)} {\Rsplit {\Rsingle{\alpha=\gamma_0,\vec\gamma}{\vec\gamma}} \alpha} \\
                         \Let{(r, c)}{\Receive c} \\
                         (r, \Rconcat\sigma{\Rsingle\alpha c} )
       \end{array}
    \\
    &\GVEReduceTo
      \begin{array}[t]{@{}l}
                          \Let{(r, c)}{\Receive {\gamma_0}} \\
                          (r, \Rconcat {\Rsingle{\vec\gamma}{\vec\gamma}}{\Rsingle\alpha c} )
      \end{array}
    \\
    &\GVEReduceTo[\gamma_0?\Trans{v'}]
                          (\Trans{v'}, \Rconcat {\Rsingle{\vec\gamma}{\vec\gamma}}{\Rsingle\alpha
      {\gamma_0}} )
    \\
    &\GVEReduceTo
      (\Trans{v'}, {\Rsingle{\alpha=\gamma_0,\vec\gamma}{\vec\gamma}} )
    \\
    &= \TransE[{\Rsingle{\alpha=\gamma_0,\vec\gamma}{\vec\gamma}} ]{v'}
  \end{align*}
  The reductions involving $\Send v v'$, $\Accept n$, $\Request n$, and $\Close n$
  are similar.

  If $t = {\Fork  t_1;t_2}$, then $\Thread t \VGRPReduceTo[\KWFork]
  \Thread{t_1} \| \Thread{t_2}$.
  \begin{align*}
    \Thread{\TransE[{\Rsingle{\vec\gamma}{\vec\gamma}}]{\Fork t_1;t_2}}
    &=
      \ThreadOpen\begin{array}[t]{l}
                                        \Let{(\sigma_1, \sigma_2)} {\Rsplitmany {\Rsingle{\vec\gamma}{\vec\gamma}}
                                        {\Dom{\Sigma_1\setminus\Sigma_2}}}
                                      \\
                                        \Let \_ {\Fork\TransE[\sigma_1]{t_1}}
                                        \\
                   \TransE[\sigma_2]{t_2}
                   \ThreadClose
                                      \end{array}
    \\
    &\GVPReduceTo
      \ThreadOpen\begin{array}[t]{l}
                                        \Let \_ {\Fork\TransE[{\Rsingle{\vec\gamma_1}{\vec\gamma_1}}]{t_1}}
                                        \\
                   \TransE[{\Rsingle{\vec\gamma_2}{\vec\gamma_2}}]{t_2}
                   \ThreadClose
                 \end{array}
    \\
    &\GVPReduceTo[\KWFork]
      \Thread {\TransE[{\Rsingle{\vec\gamma_1}{\vec\gamma_1}}]{t_1}} \|
      \Thread {\TransE[{\Rsingle{\vec\gamma_2}{\vec\gamma_2}}]{t_2}}
    \\
    \Trans{\Thread{t_1} \| \Thread{t_2}}
    &=\Trans{\Thread{t_1}} \| \Trans{\Thread{t_2}}
    \\
    &= \Thread{\Let\sigma {\Rsingle{\vec\gamma_1}{\vec\gamma_1}}{ \TransE {t_1}}} \|
      \Thread{\Let\sigma {\Rsingle{\vec\gamma_2}{\vec\gamma_2}}{ \TransE {t_2}}}
    \\
    &= \Thread{\TransE[{\Rsingle{\vec\gamma_1}{\vec\gamma_1}}] {t_1}} \|
      \Thread{\TransE[{\Rsingle{\vec\gamma_2}{\vec\gamma_2}}] {t_2}}
  \end{align*}

  Finally, consider $\Thread{ E[\New S]} \VGRPReduceTo[\KWNew]
  \Cnewap{n : S}\Thread{E[n]}$.
  \begin{align*}
    \Trans{\Thread{E[\New S]}}
    &= \Thread{\Let\sigma{\Rsingle{\vec\gamma}{\vec\gamma}} \TransE
      {E[\New S]}}
    \\
    &= \Thread{\Let\sigma{\Rsingle{\vec\gamma}{\vec\gamma}}
      \Trans{E}[\TransE{\New S}]}
    \\
    &= \Thread{\Let\sigma{\Rsingle{\vec\gamma}{\vec\gamma}}
      \Trans{E}[{({\New {\Trans S}}, \sigma)}]}
    \\
    &\GVPReduceTo \Thread{
       \Trans{E}[ {({\New {\Trans S}},
       {\Rsingle{\vec\gamma}{\vec\gamma}})}]}
    \\
    &\GVPReduceTo[\KWNew]
      \Cnewap {n:\Trans S}
      \Thread{
      \Trans{E}[{(n,
      {\Rsingle{\vec\gamma}{\vec\gamma}})}]}
    \\
    &=
      \Cnewap {n:\Trans S}
      \Thread{
      \TransE[{\Rsingle{\vec\gamma}{\vec\gamma}}]{E[n]}
      }
    \\
    \Trans{\Cnewap{n : S}\Thread{E[n]}}
    &= \Cnewap{n:\Trans S} \Trans{\Thread{E[n]}}
    \\
    &= \Cnewap{n:\Trans S}
      {\Thread{\Let\sigma{\Rsingle{\vec\gamma}{\vec\gamma}}
      \TransE{E[n]}}}
    \\
    &\GVEReduceTo
      \Cnewap {n:\Trans S}
      \Thread{
      \TransE[{\Rsingle{\vec\gamma}{\vec\gamma}}]{E[n]}
      }
      \qedhere
  \end{align*}
\end{proof}

\subsubsection{Proof of \Cref{lemma:anf-simulation}}
\label{sec:proof-prop-anf-simulation}

\begin{proof}
  \textbf{Part 1 expression reduction}

  \textbf{Suppose that }$E[e_1] \GVEReduceTo E[e_2]$. For $E=\Hole$,
  there are two cases.

  \textbf{Case }${(\Lam x e)\,v \GVEReduceTo e[v/x]}$.

  \begin{align*}
    \TOANF{(\Lam x e)\,v}
    & = {({\Lam x \TOANF{e}})\, \TOANF{v}} & \text{by value preservation}\\
    & \VGREReduceTo \TOANF{e}[ \TOANF{v} / x] & \text{substitution preservation}\\
    & = \TOANF{e[v/x]}
  \end{align*}

  \textbf{Case }${\Let{(x,y)}{(v,w)} e \GVEReduceTo e[v,w/x,y]}$.
  \begin{align*}
    \TOANF {\Let{(x,y)}{(v,w)} e}
    & = \Let{(x,y)}{(\TOANF v, \TOANF w)} \TOANF e \\
    & \VGREReduceTo \TOANF e[\TOANF v, \TOANF w / x, y] & \text{by
                                                          substitution preservation}\\
    & = \TOANF{e [v, w / x, y]}
  \end{align*}

  \textbf{Case }$E= E'\, e$. As $E[e_1]$ reduces, it must be that
  $E'[e_1] =: n$ is a non-value such that $n \GVEReduceTo e'$.
  \begin{align*}
    \TOANF{E'[e_1]\, e}
    &= \Let x {\TOANF{E'[e_1]}} \TOANF{x\,e} \\
    &\GVEReduceStar \Let x {\TOANF{e'}} \TOANF{x\,e}  & \text{by IH}\\
    &\dots = \TOANF{e'\,e} & \text{if $e'$ is a non-value} \\
    &\dots \GVEReduceTo \TOANF{x\,e}[\TOANF{e'}/x] & \text{if $e'$ is
                                                     a value} \\
    &= \TOANF{e'\,e} & \text{by substitution preservation}
  \end{align*}

  \textbf{Case }$E= v\,E'$. As $E[e_1]$ reduces, it must be that
  $E'[e_1] =: n$ is a non-value such that $n \GVEReduceTo e'$.
  \begin{align*}
    \TOANF{v\, E'[e_1]}
    &= \Let x {\TOANF{E'[e_1]}} \TOANF{v}\,x \\
    &\GVEReduceStar\Let x{\TOANF{e'}} \TOANF{v}\,x & \text{by IH}\\
    &\dots = \TOANF{v\,e'} & \text{if $e'$ is a non-value} \\
    &\dots \GVEReduceTo \TOANF{v}\,{\TOANF{e'}} & \text{if $e'$ is a
                                                  value} \\
    & \TOANF{v\, e'}
  \end{align*}

  \textbf{The remaining cases} are similar.
\end{proof}

\subsubsection{Proof of \Cref{lemma:back-simulation}}
\label{sec:proof-prop-backwards-simulation}

\begin{proof}
  \textbf{Part 1 expression reduction}

  \textbf{Case }${(\Lam x e)\,v \GVEReduceTo e[v/x]}$.

  \begin{align*}
    \Back{(\Lam x e)\,v}
    & = {({\Lam x \Back{e}})\, \Back{v}} & \text{by value preservation}\\
    & \VGREReduceTo \Back{e}[ \Back{v} / x] & \text{by substitution preservation}\\
    & = \Back{e[v/x]}
  \end{align*}

  \textbf{Case }${\Let{(x,y)}{(v,w)} e \GVEReduceTo e[v,w/x,y]}$.
\begin{align*}
    \Back {\Let{(x,y)}{(v,w)} e}
    & = \Let{(x,y)}{(\Back v, \Back w)} \Back e \\
    & \VGREReduceTo \Back e[\Back v, \Back w / x, y] & \text{by
                                                       substitution preservation}\\
    & = \Back{e [v, w / x, y]}
  \end{align*}

  \textbf{Case }$\Let x v e \GVEReduceTo e[v/x]$.

  \begin{align*}
    \Back{\Let x v e}
    &= \Let x {\Back v} \Back{e} \\
    &\VGREReduceTo \Back{e}[\Back v/x] \\
    &= \Back{e[v/x]}
  \end{align*}

  \textbf{Case }$\Let x n e \GVEReduceTo \Let x {e'} e$ because $n \GVEReduceTo e'$.
  \begin{align*}
    \Back{\Let x n e}
    &= \Let x{\Back{n}}  {\Back{e}} \\
    &\VGREReduceTo \Let x{\Back{e'}}  {\Back{ e}} & \text{by IH} \\
    &= \Back{\Let x {e'} e}
  \end{align*}

\textbf{Part 2 process reduction}

\textbf{Case }$\Thread{E[\Fork e]} \GVPReduceTo[\KWFork] \Thread{E[\UnitV]}
\| \Thread e$.
\begin{align*}
  \Back{\Thread{E[\Fork e]}}
  &= \Thread{\Back E[\Fork{\Back e};\UnitV]} \\
  &\VGRPReduceTo[\KWFork] \Thread{\Back E[\UnitV]} \| \Thread{{\Back
    e}}\\
  &= \Back{\Thread{E[\UnitV]} \| \Thread e}
\end{align*}

\textbf{Case }$\Thread{E[\New s]} \GVPReduceTo[\KWNew] (\nu p)
\Thread{E[p]}$.
\begin{align*}
  \Back{\Thread{E[\New s]}}
  &= \Thread{\Back E[\New{\Back s}]} \\
  &\VGRPReduceTo[\KWNew] (\nu p:\Ap{\Back s}) \Thread{\Back E[p]} \\
  &= \Back{(\nu p)~\Thread{E[p]}}
\end{align*}

\textbf{Case }$\Thread{E[\Accept p]} \| \Thread{F[\Request p]}
\GVPReduceTo[\KWAccept]
(\nu\gamma\delta) (\Thread{E[\gamma]} \| \Thread{F[\delta]})$.
\begin{align*}
  \Back{\Thread{E[\Accept p]} \| \Thread{F[\Request p]}}
  &={\Thread{\Back E[\Accept p]} \| \Thread{\Back F[\Request p]}} \\
  &\VGRPReduceTo[\KWAccept] (\nu\gamma) {\Thread{\Back E[\gamma^+]} \|
    \Thread{\Back F[\gamma^-]}} \\
  &= \Back{(\nu\gamma^+\gamma^-)~ \Thread{E[\gamma^+]} \| \Thread{F[\gamma^-]}}
\end{align*}

\textbf{Case }$(\nu\gamma^+\gamma^-) (\Thread{E[\Send v {\gamma^+}]} \| \Thread{F[\Receive {\gamma^-}]})
\GVPReduceTo[\KWSend]
(\nu\gamma^+\gamma^-) (\Thread{E[\gamma^+]} \| \Thread{F[(v, \gamma^-)]})
$.
\begin{align*}
  \qquad &
  \Back{(\nu\gamma^+\gamma^-) (\Thread{E[\Send v {\gamma^+}]} \|
  \Thread{F[\Receive {\gamma^-}]})} \\
  &= (\nu \gamma)~
    \Thread{\Back E[\Let z {\Send{\Back v}{\gamma^+}} {\gamma^+}]} \|
    \Thread{\Back F[\Let x{ \Receive{\gamma^-}}{(x, \gamma^-)}]} \\
  &\VGRPReduceTo[\KWSend]
    (\nu \gamma)~
    \Thread{\Back E[\Let z {\UnitV} {\gamma^+}]} \|
    \Thread{\Back F[\Let x{ \Back v}{(x, \gamma^-)}]} \\
  &\VGRPReduceTo
    (\nu \gamma)~
    \Thread{\Back E[{\gamma^+}]} \|
    \Thread{\Back F[\Let x{ \Back v}{(x, \gamma^-)}]} \\
  &\VGRPReduceTo
    (\nu \gamma)~
    \Thread{\Back E[{\gamma^+}]} \|
    \Thread{\Back F[{(\Back v, \gamma^-)}]} \\
  & = \Back{(\nu\gamma^+\gamma^-)~
    \Thread{E[\gamma^+]} \} \Thread{F[(v, \gamma^-)]}
    }
    \qedhere
\end{align*}
\end{proof}

\subsubsection{Proof of \Cref{lemma:anf-compatible}}
\label{sec:proof-lemma-anf-compatible}

\begin{proof}
  The proof is by induction on the derivation of  $\JGVExpr[']\Gamma e
  {t / \Sigma \mapsto \Sigma'}$.

  \textbf{Case }\TirName{T-Unit'}, \TirName{T-Var'},
  \TirName{T-LamU'}, \TirName{T-LamL'}: all immediate by IH\@.

  \textbf{Case }$\ruleGVappEFF$. By IH, we have that
  \begin{gather}
    \label{eq:7}
    \JGVExpr[']{\Gamma_1}{\TOANF {e_1}}{
      t_2 {\linto}^{\Sigma_2\mapsto\Sigma_3} t_1 / \Sigma_0\mapsto\Sigma_1} \\
    \label{eq:8}
    \JGVExpr[']{\Gamma_2}{\TOANF{e_2}}{t_2/\Sigma_1\mapsto\Sigma_2,\Sigma_2'}
  \end{gather}
  There are four subcases.

  \textbf{Subcase }$e_1, e_2$ are non-values. Then $\TOANF{e_1\,e_2}= \Let
  x{\TOANF{e_1}} \Let y {\TOANF{e_2}}{x\,y}$.
  Let $\Gamma_3 =
  x: t_2 {\linto}^{\Sigma_2\mapsto\Sigma_3} t_1,
  y :  t_2$. By \TirName{T-App'} we obtain
  \begin{gather}
    \label{eq:9}
    \JGVExpr[']{\Gamma_3}{x\,y}{t_1 / \Sigma_2,\Sigma_2' \mapsto
      \Sigma_3, \Sigma_2'}
  \end{gather}
  By \TirName{T-Let'} using the obvious splitting $
  \Gamma_2,   x: t_2 {\linto}^{\Sigma_2\mapsto\Sigma_3} t_1
  = \Gamma_2 +   [x: t_2
  {\linto}^{\Sigma_2\mapsto\Sigma_3} t_1] $ we obtain
  \begin{gather}
    \label{eq:10}
    \JGVExpr[']{\Gamma_2,   x: t_2 {\linto}^{\Sigma_2\mapsto\Sigma_3} t_1}{\Let y {\TOANF{e_2}} x\,y}{t_1 /\Sigma_1
      \mapsto \Sigma_3,\Sigma_2'
    }
  \end{gather}
  By \TirName{T-Let'} using the splitting $\Gamma =
  \Gamma_1+\Gamma_2$,~\eqref{eq:7}, and~\eqref{eq:10} we obtain
  \begin{gather}
    \label{eq:11}
    \JGVExpr[']{\Gamma}{\Let x{\TOANF{e_1}}\Let y {\TOANF{e_2}} x\,y}{
      t_1 / \Sigma_0\mapsto\Sigma_3,\Sigma_2'
    }
  \end{gather}

  \textbf{Subcase }$e_1, e_2$ are values. Then $\TOANF{e_1\,e_2} =
  \TOANF{e_1}\,\TOANF{e_2}$. In that case, $\Sigma_0 = \Sigma_1 =
  \Sigma_2,\Sigma_2'$ and we can apply \TirName{T-App'} directly to
  the IH~\eqref{eq:7} and~\eqref{eq:8} to get
  \begin{gather}
    \label{eq:14}
    \JGVExpr\Gamma{\TOANF{e_1}\,\TOANF{e_2}}{t_1 / \Sigma_2, \Sigma_2'
      \mapsto \Sigma_3, \Sigma_2'}
  \end{gather}

  \textbf{Subcase } mixed cases: The same principles apply.

  \textbf{Case} the remaining cases are similar.
\end{proof}

\subsubsection{Proof of
  \Cref{proposition:typing-preservation-backwards}}
\label{sec:proof-prop-typing-preservation-backwards}
\begin{proof}
  We need an auxiliary statement about VGR value typing that is proved by
  mutual induction with the main statement.

  \begin{itemize}
  \item Suppose that $\JGVExpr[']\Gamma v {t / \Sigma_1\mapsto
      \Sigma_2}$.
      Then $\Sigma_1 = \Sigma_2$ can be chosen arbitrarily and
      $\JVGRValue {\Back\Gamma} {\Back v} {\Back t}$ is a value typing
      in VGR\@.
  \end{itemize}

  \textbf{Case }$\ruleGVvarEFF$.
  Immediate by \TirName{C-Var}:  $\JVGRValue{\Back\Gamma, x:
    \Back t}x {\Back t}$.

  \textbf{Case }$\ruleGVunitEFF$.
  Immediate by \TirName{C-Const}: $\JVGRValue{\Back\Gamma}\UnitV\TUnit$.

  \textbf{Case }$\ruleGVlamlinEFF$.

  By induction on the main statement we have
  for all $\Sigma$ with $\Sigma\#\Sigma_0$ and $\Sigma\#\Sigma_1$:
  \begin{gather*}
    \JVGRExpr {\Back{\Gamma, x:t_2}} {\Sigma,
      \Sigma_0} {\Back e} {\Back {t_1}} {\Sigma} {\Sigma_1}
  \end{gather*}
  Choosing $\Sigma=\emptyset$ we obtain  by \TirName{C-Abs}
  \begin{gather*}
    \JVGRValue{\Back\Gamma}{\Back{\Lam x e}}{\Sigma_0
      ; \Back{t_2} \to \Back{t_1} ; \Sigma_1}
    \intertext{hence}
    \JVGRValue{\Back\Gamma}{\Back{\Lam x e}}{
       \Back{t_2 \to^{\Sigma_0\mapsto\Sigma_1} t_1}}
  \end{gather*}
  \begin{itemize}
  \item For the main claim suppose now that $\JGVExprID\Gamma e {t / \Sigma_0\mapsto\Sigma_1}$.
  \end{itemize}

  \textbf{Case }$e$ is a value $v$. Hence,
  $\JVGRValue {\Back\Gamma} {\Back v} {\Back t}$ and
  $\Sigma_0=\Sigma_1$ by IH\@.
  Moreover, the choice of $\Sigma_0$ is arbitrary.
  By rule \TirName{C-Val}, we obtain that for all $\Sigma$,
  $$\JVGRExpr{\Back\Gamma} \Sigma v {\Sigma} {\Back t} {\emptyset}$$

  \textbf{Case }$\ruleGVappEFF$.

  As $e_1\, e_2$ is an application in LFST-ANF,  $e_1$ and $e_2$ are both values.
  By IH, $\Sigma_0=\Sigma_1 = \Sigma_2,\Sigma_2'$ and
  \begin{align*}
    \JVGRValue {\Back{\Gamma_1}} {\Back {e_1}} {\Back{t_2
        \linto^{\Sigma_2\mapsto\Sigma_3} t_1}}
    & =
    \JVGRValue {\Back{\Gamma_1}} {\Back {e_1}} {\Sigma_2;\Back{t_2}
      \to \Back{ t_1}; \Sigma_3}
    \\
    \JVGRValue {\Back{\Gamma_2}} {\Back {e_2}} {\Back{t_2}}
  \end{align*}
  By weakening
  \begin{gather*}
    \JVGRValue {\Back{\Gamma}} {\Back {e_1}} {\Sigma_2;\Back{t_2}
      \to \Back{ t_1}; \Sigma_3}
    \\
    \JVGRValue {\Back{\Gamma}} {\Back {e_2}} {\Back{t_2}}
  \end{gather*}
  Applying \TirName{C-App} yields
  \begin{gather*}
    \JVGRExpr\Gamma {\Sigma_2,\Sigma_2'}{\Back{e_1}\,\Back{e_2}}{\Sigma_2'}{\Back{t_1}}{\Sigma_3}
  \end{gather*}
  The claim follows because $\Sigma_2'$ is arbitrary.

  \textbf{Case }$\ruleGVletEFF$.

  By IH we have that, for all $\Sigma_0$ with $\Sigma_0\#\Sigma$ and
  $\Sigma_0\#\Sigma'$ and $\Sigma_0\#\Sigma''$,
  \begin{gather}
    \label{eq:15}
    \JVGRExpr {\Back{\Gamma_1}} {\Sigma_0, \Sigma} {\Back {e_1}}
    {\Back {t_1}} {\Sigma_0} {\Sigma'}
    \intertext{and}
    \label{eq:16}
    \JVGRExpr {\Back{\Gamma_2, x:t_1}} {\Sigma_0, \Sigma'} {\Back {e_2}}
    {\Back {t_2}} {\Sigma_0} {\Sigma''}
  \end{gather}
  Observe that $\Sigma_0\cap\Sigma_0 = \Sigma_0$ and $\Sigma_0\cap
  \Sigma' = \emptyset$ by assumption. Hence, \TirName{C-Let} is
  applicable and yields
  \begin{gather}
    \label{eq:17}
    \JVGRExpr\Gamma {\Sigma_0,\Sigma}{\Let x {\Back{e_1}} {\Back{e_2}}}{\Sigma_0}{\Back{t_1}}{\Sigma''}
  \end{gather}

  \textbf{Case }$\ruleGVforkEFF$.

  By IH we have that, for all $\Sigma_0$ with $\Sigma_0\#\Sigma$,
  \begin{gather}
    \label{eq:18}
    \JVGRExpr\Gamma {\Sigma_0,\Sigma}{ {\Back{e}}}{\Sigma_0}{\Back{t}}{\emptyset}
  \end{gather}
  Now $\Back{\Fork e} = \Fork {\Back e}; ()$ and, for all $\Sigma_0$,
  \begin{gather}
    \label{eq:19}
    \JVGRExpr\Gamma{\Sigma_0}{()}{\Sigma_0}{\TUnit}{\emptyset}
  \end{gather}
  Applying \TirName{C-Fork} to~\eqref{eq:18} and~\eqref{eq:19} yields
  \begin{gather}
    \label{eq:20}
    \JVGRExpr\Gamma{\Sigma_0,\Sigma}{\Fork{\Back e}; ()}{\Sigma_0}{\TUnit}{\emptyset}
  \end{gather}
  for all $\Sigma_0\#\Sigma$.

  \textbf{Case }$\ruleGVsendEFF$.

  Recall that $e_1$ and $e_2$ are values due to LFST-ANF\@.
  Hence $\Sigma=\Sigma' = \Sigma'',\alpha: \Outp t s$.
\end{proof}

\end{document}